\documentclass[reprint,twocolumn,prx,amsmath,amssymb,superscriptaddress]{revtex4}
\usepackage{amsmath}
\usepackage{amssymb}
\usepackage{graphicx}
\usepackage{color}
\usepackage{amsfonts}
\usepackage{subfigure}
\usepackage{enumitem}
\begin{document}

\title{The Master Stability Function for Synchronization in Simplicial Complexes}

\author{L. V. Gambuzza$^{*}$}
\affiliation{Department of Electrical, Electronics and Computer Science Engineering, University of Catania, 95125 Catania, Italy}
\author{F. Di Patti$^{*}$}
\affiliation{CNR-Institute of Complex Systems, Via Madonna del Piano, Sesto Fiorentino, Florence 50019, Italy}
\author{L. Gallo$^{*}$}
\affiliation{Department of Physics and Astronomy,  University of Catania, 95125 Catania, Italy}
\author{S. Lepri}
\affiliation{CNR-Institute of Complex Systems, Via Madonna del Piano, Sesto Fiorentino, Florence 50019, Italy}
\author{M. Romance}
\affiliation{Department of Applied Math. and Data, Complex Networks and Cybersecurity Research Institute, University Rey Juan Carlos, C/Tulipán s/n, Móstoles, Madrid 28933, Spain}
\author{R. Criado}
\affiliation{Department of Applied Math. and Data, Complex Networks and Cybersecurity Research Institute, University Rey Juan Carlos, C/Tulipán s/n, Móstoles, Madrid 28933, Spain}
\author{M. Frasca$^{+}$}
\affiliation{Department of Electrical, Electronics and Computer Science Engineering, University of Catania, 95125 Catania, Italy}
\author{V. Latora$^{+}$}
\affiliation{Department of Physics and Astronomy,  University of Catania, and INFN, 95125 Catania, Italy}
\affiliation{School of Mathematical Sciences, Queen Mary University of London, London E1 4NS, UK}
\affiliation{The Alan Turing Institute, The British Library, London NW1 2DB, United Kingdom}
\author{S. Boccaletti$^{+}$}
\affiliation{CNR-Institute of Complex Systems, Via Madonna del Piano, Sesto Fiorentino, Florence 50019, Italy}
\affiliation{Unmanned Systems Research Institute, Northwestern Polytechnical  University, Xi'an 710072, China}
\affiliation{Moscow Institute of Physics and Technology, Dolgoprudny, Moscow Region, 141701, Russian Federation}

%\author{L. V. Gambuzza$^{a,*}$, F. di Patti$^{b,*}$, L. Gallo$^{c,*}$, S. Lepri$^b$, M. Romance$^d$, R. Criado$^d$,
%M. Frasca$^{a,+}$, V. Latora$^{c,e,+}$, and S. Boccaletti$^{b,f,g,+}$}
%\affiliation{$^a$ Department of Electrical, Electronics and Computer Science Engineering, University of Catania, 95125 Catania, Italy \\
%$^b$ CNR-Institute of Complex Systems, Via Madonna del Piano, Sesto Fiorentino, Florence 50019, Italy \\
%$^c$ Department of Physics and Astronomy,  University of Catania, 95125 Catania, Italy\\
%$^d$ Department of Applied Math. and Data, Complex Networks and Cybersecurity Research Institute, University Rey Juan Carlos, C/Tulipán s/n, Móstoles, Madrid 28933, Spain\\
%$^e$ School of Mathematical Sciences, Queen Mary University of London, London E1 4NS, UK \\
%$^f$ Unmanned Systems Research Institute, Northwestern Polytechnical  University, Xi'an 710072, China \\
%$^g$ Moscow Institute of Physics and Technology, Dolgoprudny, Moscow Region, 141701, Russian Federation \\

\date{\today}

\begin{abstract}
All interesting and fascinating collective properties of a complex system arise from the intricate way in which its components interact.  Various systems in physics, biology, social sciences and engineering have been successfully modelled as networks of coupled dynamical systems, where the graph links describe pairwise interactions. This is, however, too strong a limitation, as recent studies have revealed that higher-order many-body interactions are present in social groups, ecosystems and in the human brain, and they actually affect the emergent dynamics of all these systems.
Here, we introduce a general framework that allows to study coupled dynamical systems accounting for the precise microscopic structure of their interactions at
any possible order. We consider the most general ensemble of identical dynamical systems, organized on the nodes of a simplicial complex, and interacting through synchronization-non-invasive coupling function. The simplicial complex can be of any dimension, meaning that it can account, at the same time, for pairwise interactions, three-body interactions and so on.
In such a broad context, we show that complete synchronization, a circumstance where all the dynamical units arrange their evolution in unison, exists as an invariant solution, and we give the necessary condition for it to be observed as a stable state in terms of a Master Stability Function.
This generalizes the existing results valid
for pairwise interactions (i.e. graphs) to the case of complex systems with the most general possible architecture.
Moreover, we show how the approach can be simplified for specific, yet frequently occurring, instances, and we verify all our theoretical predictions in synthetic and real-world systems.  Given the completely general character of the method proposed, our results contribute to the theory of dynamical systems with many-body interactions and can find applications in an extremely wide range of practical cases.
\end{abstract}

\maketitle

$^*$ These Authors contributed equally to this work

$^+$ These Authors contributed equally to this work

%PRX Subject Areas: Complex Systems, Nonlinear Dynamics

\section{Introduction}

Many systems in physics, biology, engineering and social sciences can be modeled as networks of interacting units \cite{boccaletti2006complex}.
Often, each of the elementary system constituents (the nodes of the network) is a dynamical system itself, whose evolution is influenced by the states of the other units to which is connected to through the links of the network. Unravelling how the interplay of network structure and the type of interactions shape the overall dynamics of the system and rule its collective
behaviors is thus a problem of wide interest across disciplines.

There is, however, an underlying strong assumption that is made when one adopts a network representation of a complex system:
the overall interplay among the unitary components of the system
is assumed to be exhaustively described by combinations of pairwise interactions. Such an hypothesis may be justified when studying certain types of processes, but it is very short in representing faithfully other many circumstances. Indeed, from functional
\cite{petri2014homological, lord2016insights, lee2012persistent} and
structural \cite{sizemore2018cliques} brain networks to protein
interaction networks \cite{estrada2018centralities}, to semantic networks \cite{sizemore2017knowledge} and co-authorship graphs in science \cite{patania2017shape} there are a lot of practical situations which simply cannot be factorized in terms of pairwise interactions \cite{petri2013topological, sizemore2016classification}.

Simplicial complexes (SCs) are topological structures formed by simplices of different dimensions (such as nodes, links, triangles, tetrahedra, etc..) and map many-body interactions between the elements of a system.
Differently from networks, SCs can therefore efficiently represent the interactions between any number of units. While SCs are not a new idea \cite{aleksandrov1998combinatorial}, the availability of new datasets and the recent advances in
topological data analysis techniques \cite{carlsson2009topology} renewed the interest of the scientific community \cite{salnikov2018simplicial, sizemore2018importance}.
In particular, a lot of attention
in the last years has been devoted to the modelling of simplicial complexes, and significant progresses were made in extending to SCs standard graph models, such as random graphs models \cite{costa2016random}, the configuration model \cite{courtney2016generalized}, models of network growth \cite{bianconi2015complexa} and activity driven models \cite{petri2018simplicial}.

On the other hand, synchronization is a phenomenon appearing ubiquitously in natural and engineered systems \cite{pikovsky2003synchronization,boccaletti2018synchronization}, and corresponds to the emergence of a collective behavior wherein the system unitary components eventually adjust themselves into a common evolution in time. Various studies have
shed light on the intimate relationships between the topology of a networked system, its synchronizability, and the properties of the synchronized states.
In particular, synchronous behaviors have been observed and characterized in small-world ~\cite{barahona2002synchronization}, weighted \cite{chavez2005weighted},  multilayer~\cite{delgenio2016layers}, and adaptive networks \cite{gutierrez2011emerging,avalos2018adaptive}.
Outside complete synchronization, moreover, other types of synchronization have been revealed to emerge in networked systems,  including remote synchronization~\cite{gambuzza2013analysis,nicosia2013remote}, cluster states~\cite{pecora2014cluster} and synchronization of group of nodes~\cite{gambuzza2018distributed}, chimera~\cite{abrams2004chimera,panaggio2015chimera} and
Bellerophon states \cite{bi2016coexistence,xu2018origin}.
Finally, the transition to synchronization has been shown to be either smooth and reversible, or abrupt and irreversible (as in the case of explosive synchronization, resembling a first-order like phase transition~\cite{boccaletti2016explosive}).

While attempts of extending to $p-$uniform hypergraphs the analysis of complete synchronization of dynamical systems have been recently made \cite{krawiecki2014chaotic}, the study of systems interplaying through higher order interactions in simplicial complexes has been so far limited to the case of the Kuramoto model \cite{acebron2005kuramoto,rodrigues2016kuramoto}. This is, in fact, a  specific model, wherein each unit of the ensemble $i=1, ..., N$ is a phase oscillator and is characterized by the evolution of its real valued phase $\theta_i(t) \in [0,2\pi]$.
The model has been studied in all different sorts of network's topologies with possible applications to biological and social systems \cite{acebron2005kuramoto,boccaletti2018synchronization}, and recently extensions of it have been proposed that include higher-order interactions. Namely, it has been shown that the Kuramoto model may exhibit abrupt desynchronization when three-body interactions among all the oscillators are added to \cite{tanaka2011multistable}, or completely replace \cite{skardal2019abrupt}, the all-to-all pairwise interactions of the original model.
Similar results have been obtained with a  non-symmetric variation of the Kuramoto model in which the microscopic details of the interactions among the phase oscillators are described in the form of a simplicial complex \cite{skardal2019higher}.

A different approach has been proposed by Mill\'an et al,
who have formulated a higher-order Kuramoto model in which
the oscillators are placed not on the nodes but on higher-order simplices, such as links, triangles, and so on, of a simplicial complex \cite{millan2019explosive}.
Finally, Lucas et al. have considered an extension of the Kuramoto model to high order interactions of any order, which is still analytically tractable because all the oscillators have identical frequencies  \cite{maxime2020arxiv}.

We here abandon the limitation of sticking with a specific model system, and introduce instead the most general framework for the study of dynamical systems in SCs. Namely, we consider an ensemble of completely generic (yet identical) dynamical systems, organized on the nodes of a simplicial complex of generic order, and interacting via generic coupling functions. In other words, except for the fact that the systems have to be identical, we {\it do not make} any specific assumption that may limit in a way or another our approach.
In such a wide context, we show that complete synchronization exists as an invariant solution as far as the coupling functions cancel out when nodes dynamics is identical. Furthermore, we give the necessary condition for it to be observed as a stable state in terms of a Master Stability Function, a method initially developed in Ref. \cite{pecora1998master} for pairwise coupled systems, and later extended in many ways to complex networks \cite{sun2009master} and to time-varying interactions \cite{stilwell2006sufficient,frasca2008synchronization,zhou2016synchronization}.
Therefore, not only our framework includes and encompasses all studies made so far on the Kuramoto model, but it is valid for an enormously larger number of situations, and as so it is applicable to a very wide range of experimental and/or practical circumstances. We will show, indeed, that all the theoretical predictions that our method entitles us to make are fully verified in simulations of synthetic and real-words networked systems.

\section{Networks and higher-order structures}

A network is a collection of nodes
and of edges connecting pairs of
nodes. Mathematically, it is represented by a graph
$\mathcal{G}  =(\mathcal{V},\mathcal{E})$,
which consists of a set $\mathcal{V}$ with $N=|\mathcal{V}|$ elements called vertices (or nodes), and a set $\mathcal{E}$ whose $K$ elements, called edges or links, are pairs of
nodes $(i,j)$ ($i,j=1,2,\ldots,N$ and $i \neq j$).
As graphs explicitly refer to pairwise interactions, networks have been very successful in
capturing the properties of coupled dynamical systems in all such cases in which the interactions can be expressed (or approximated) as a sum of two-body terms~\cite{latora_nicosia_russo_2017}.
Conversely, their limits emerge when it comes to model higher-order interactions.
In fact, the presence of a triangle of three nodes $i, j, k$
in a network, e.g. the presence of the
three links $(i, j)$, $(i, k)$, $(j, k)$ in the corresponding graph, is not able to capture the difference between a three-body interaction of the three individuals, from the sum of three pairwise interactions. Notice that these are two completely different situations, with completely different social mechanisms and dynamics at work
\cite{iacopini2019simplicial}.

Simplicial complexes are instead the proper mathematical structures for describing high order interactions. A simplicial complex is an aggregate of simplices, objects that generalize links and can in general be of different dimension.
A {\em $d$-simplex}, or simplex of dimension $d$,  $\sigma$ is, in its simplest definition, a collection of $d+1$ nodes. In this way, a 0-simplex is a node, a 1-simplex is a link, a 2-simplex $(i,j,k)$ is a two-dimensional object made by three nodes, usually called a (full) triangle, a 3-simplex is a tetrahedron, i.e. a three-dimensional object and so on.
It is now possible to differentiate
between a three-body interaction, and three bodies in pairwise interactions: the first case will be represented by a  complete triangle, a 2-dimensional simplex,
while the second case will consist of three 1-dimensional objects. Hence, in the following of this paper, simplices of dimension $d$ will be used to describe the structure of $(d+1)$-body interactions.

Finally, a {\em simplicial complex} $\cal S$ on a given set of nodes $\mathcal{V}$, with
$|\mathcal{V}|=N$, is a collection of $M$ simplices,  ${\cal S} = \{ \sigma_1, \sigma_2,
\ldots, \sigma_M \}$, with the extra requirement that, for any simplex $\sigma \in \cal S$, all the simplices $\sigma'$ with $\sigma' \subset \sigma$, i.e. all the simplices built from subsets of $\sigma$, are also contained in $\cal S$. Due to this requirement, SCs are a very particular type of hypergraphs \cite{berge1973graphs}.
SCs have shown to be appropriate in the context of social systems \cite{kee2013social,iacopini2019simplicial,alvarez2020evolutionary} and, as we will see in the next Section, they will turn very useful to study coupled dynamical systems.
In the following, we will
indicate as $M_d$, $d=1,2,\ldots D$
the number of $d$-simplices present in $\cal S$ (where $D$, the order of the simplicial complex, is the dimension of the
largest simplex in $\cal S$), and we have the
constraint $\sum_{d=1}^D M_d = M$.

As a mathematical representation of SCs, we will use here a formalism which   generalises  directly the concept of adjacency matrix for a network.
The adjacency matrix $A$ of a graph ${\cal G}$ is a  $N \times N$ matrix, such that entry  $a_{ij}$ is 1 when edge $(i,j) \in {\cal E}$, and 0 otherwise. The idea can be extended to SCs by considering tensors instead of matrices.
In fact, for each dimension $d$, we can define the
$\underbrace{N\times N \times \dots \times N}_d$ {\em adjacency tensor} $\mathrm{A}^{(d)}$, whose entry $a^{(d)}_{i_1,\dots,i_d}$ is equal to 1 if the $d$-simplex $(i_1,\dots,i_d)$ belongs to the simplex ${\cal S}$, and is 0 otherwise~\citep{courtney2016generalized}.
Notice that each tensor is symmetric with respect to its $d$ indices, which means that the value of a given entry $a^{(d)}_{i_1,\dots,i_d}$ is equal to the value of the entries corresponding to any permutation of the indices.

With the definition above, $\mathrm{A}^{(1)}$ coincides with the standard adjacency matrix $A$, while the $N \times N \times N$ adjacency tensor
$\mathrm{A}^{(2)}$ characterizes two-dimensional objects: one has  $a_{ijk}^{(2)}=1$ if the three nodes $i$, $j$, $k$ form a full triangle,  and otherwise $a_{ijk}^{(2)}=0$.
As a conclusion, it is possible to map completely the connectivity structure of a simplicial complex
$\cal{S}$ into the entire set of $D$ adjacency tensors $\mathrm{A}^{(d)}$, $d=1,2,\ldots D$.

A node $i$ of a simplicial complex $\cal{S}$  cannot be, therefore, characterized only by giving its degree $k_i = \sum_{j=1}^N a^{(1)}_{ij}$, but one needs instead to account for the number of simplices of any dimension, incident in $i$. It is therefore extremely useful to define the generalized degree, $k_i^{(d)}$, of a node $i$ as
\begin{equation}
   k_i^{(d)} = \frac{1}{d!}\sum_{i_2=1}^N  \sum_{i_3=1}^N \ldots \sum_{i_d=1}^N a^{(d)}_{i,i_2,\ldots,i_d},
\end{equation}
with $d=1,2,\ldots,D$ so that $k_i^{(1)}$ coincides with the standard degree, $k_i^{(2)}$ counts the number of triangles (2-simplices) to which $i$ participate
$$  k_i^{(2)}= 1/2  \sum_{j=1}^N \sum_{k=1}^N a_{ijk}^{(2)},$$
$k_i^{(3)}$ the number of tetrahedrons, and so
on.

The Laplacian is a matrix that is of particular importance in many linear processes such as diffusion in graphs, but also turns useful in the linearization of nonlinear systems, for instance when we study the stability of a  synchronized state in a networked  dynamical system. The Laplacian matrix  $ \mathcal{L} = \{l_{ij}  \}$ of a graph can be defined
as $L  = K- A$, where  $K$ is the diagonal matrix having the node degrees as diagonal elements.

We here give the definition of a {\it generalized Laplacian}, describing the case of systems with high-order interactions, as the matrix $ \mathcal{L}^{(d)}$ whose elements are
\begin{equation}\label{eq:genLaplacian}
        \mathcal{L}^{(d)}_{ij} =
    \begin{cases}
    0 & \quad \text{for} \ i \neq j \ \ \  \text{and} \ \ \  a^{(1)}_{ij}=0 \\
    -k_{ij}^{(d)}  & \quad \text{ for } i \neq j \ \ \ \text{and} \ \ \ a^{(1)}_{ij}=1 \\
    d! \ k^{(d)}_i  & \quad   \text{ for } i = j,
    \end{cases}
\end{equation}
   where $k^{(d)}_{ij}$ is the generalized {\it d-degree} of the link ij, i.e. the number of (d+1)-uniform hypergraphs having the link between $i$ and $j$ as an edge, and $k^{(d)}_{i}$ is the
  generalized  {\it d-degree} of node $i$.
Notice that $\mathcal{L}^{(1)}$ recovers exactly the classical Laplacian matrix.

\section{Dynamical systems with higher-order interactions}

The object of our study is a unconditional ensemble of $N$ dynamical systems interplaying by means of $d+1$-body interactions with $d=1,\ldots,D$, whose underlying coupling structure can be therefore conveniently represented by a simplicial complex of order $D$.
The equations of motion are
\begin{equation}\label{eq:general}
\begin{array}{lll}
\dot{\mathbf{x}}_i & = & \mathbf{f}(\mathbf{x}_i ) + \sigma_{1}\sum_{j_1=1}^{N} a_{ij_1}^{(1)} \: \mathbf{g}^{(1)}(\mathbf{x}_i, \mathbf{x}_{j_1}) \\
& & + \sigma_{2} \sum_{j_1=1}^N \sum_{j_2=1}^N a_{ij_{1}j_{2}}^{(2)} \: \mathbf{g}^{(2)}(\mathbf{x}_i, \mathbf{x}_{j_1}, \mathbf{x}_{j_2})+\ldots\\
& & +
\sigma_{D} \sum_{j_1=1}^N ... \sum_{j_D=1}^N a_{ij_1....j_D}^{(D)} \: \mathbf{g}^{(D)}(\mathbf{x}_i, \mathbf{x}_{j_1}, ...,  \mathbf{x}_{j_D}),
\end{array}
\end{equation}

\noindent where $\mathbf{x}_i(t)$ is the $m$-dimensional vector state describing the dynamics of unit $i$, $\sigma_{1},..., \sigma_{D}$ are real valued parameters describing coupling strengths, $\mathbf{f}: \mathbb{R}^m \longrightarrow \mathbb{R}^m$ describes the local dynamics (which is assumed identical for all units), while $\mathbf{g}^{(d)}: \mathbb{R}^{(d+1)\times m} \longrightarrow \mathbb{R}^m$ ($d=1,....,D$) are synchronization non-invasive functions (i.e. $\mathbf{g}^{(d)}(\mathbf{x}, \mathbf{x}, ...,  \mathbf{x})  \equiv 0 \ \forall d$) ruling the interaction forms at different orders. Furthermore, for $d=1,...,D$, $a_{ij_1...j_d}^{(d)}$ are the entries of the adjacency tensor $\mathrm{A}^{(d)}$.

As notation is long and somehow cumbersome, for the sake of clarity in what follows we illustrate our study for the case of $D=2$ (so that a reader can appreciate, outside formalities, each and every conceptual action we are making), and at the end we will summarize the steps one has to do in order to extrapolate our results to all values of $D$.

Let us then consider the following set of coupled differential equations
\begin{equation}\label{eq:msf}
\begin{array}{lll}
\dot{\mathbf{x}}_i & =  & \mathbf{f}(\mathbf{x}_i ) + \sigma_{1}\sum_{j=1}^{N} a_{ij}^{(1)} \: \mathbf{g}^{(1)}(\mathbf{x}_i, \mathbf{x}_j) \\& & + \sigma_{2} \sum_{j=1}^N \sum_{k=1}^N a_{ijk}^{(2)} \: \mathbf{g}^{(2)}(\mathbf{x}_i, \mathbf{x}_j, \mathbf{x}_k),
\end{array}
\end{equation}
\noindent where $\sigma_{1}$ and $\sigma_{2}$ are the coupling strengths associated to two- and three-body interactions.

Existence and invariance of the synchronized solution $\mathbf{x}^s(t) = \mathbf{x}_1(t)  = \ldots = \mathbf{x}_N(t) $ is warranted by the non-invasiveness of the coupling functions.

\section{Master stability function}
\label{sec:generalcase}

In order to study the stability of the synchronization solution, one considers small perturbations around the synchronous state, i.e., $\delta \mathbf{x}_i = \mathbf{x}_i - \mathbf{x}^s$, and perform a linear stability analysis of Eq.  \eqref{eq:msf}. One has

\begin{equation}\label{eq:perturbations}
\begin{array}{lll}
\dot{ \delta \mathbf{x}}_i & = & J  \mathbf{f}(\mathbf{x}^s)   \delta \mathbf{x}_i + \sigma_{1} \sum_{j=1}^{N} a^{(1)}_{ij} \:
 \biggl [ \frac{\partial \mathbf{g}^{(1)}( \mathbf{x}_i, \mathbf{x}_j)}{\partial  \mathbf{x}_i }   \bigg|_{(\mathbf{x}^s, \mathbf{x}^s) } \delta \mathbf{x}_i \\
& & +  \frac{\partial  \mathbf{g}^{(1)}( \mathbf{x}_i, \mathbf{x}_j) }{\partial  \mathbf{x}_j } \bigg|_{(\mathbf{x}^s, \mathbf{x}^s) } \delta \mathbf{x}_j  \biggr ]
\\
& & + \sigma_{2}  \sum_{j=1}^N \sum_{k=1}^N a_{ijk}^{(2)} \:  \biggl [
\frac{\partial \mathbf{g}^{(2)}( \mathbf{x}_i, \mathbf{x}_j, \mathbf{x}_k)}{\partial  \mathbf{x}_i }   \bigg|_{(\mathbf{x}^s,\mathbf{x}^s, \mathbf{x}^s) } \delta \mathbf{x}_i \\
& & + \frac{\partial \mathbf{g}^{(2)}(  \mathbf{x}_i, \mathbf{x}_j, \mathbf{x}_k)}{\partial  \mathbf{x}_j }   \bigg|_{(\mathbf{x}^s,\mathbf{x}^s, \mathbf{x}^s) } \delta \mathbf{x}_j   \\
& & +\frac{\partial  \mathbf{g}^{(2)}(  \mathbf{x}_i, \mathbf{x}_j, \mathbf{x}_k) }{\partial  \mathbf{x}_k } \bigg|_{(\mathbf{x}^s,\mathbf{x}^s, \mathbf{x}^s) } \delta \mathbf{x}_k \biggr ],
\end{array}
\end{equation}

\noindent where $J  \mathbf{f}(\mathbf{x}^s) $ denotes the $m\times m$ Jacobian matrix of the function $\mathbf{f}$, evaluated at the synchronous state $\mathbf{x}^s$.

Now, we make our first, very important, conceptual step.
It consists in noticing that all coupling functions are synchronization non invasive (i.e. $\mathbf{g}^{(1)}(\mathbf{x}, \mathbf{x})  \equiv 0$ and $\mathbf{g}^{(2)}(\mathbf{x}, \mathbf{x},\mathbf{x}) \equiv 0$). As their value is then constant (equal to zero) at the synchronization manifold, it immediately follows that their total derivative vanishes as well, which implies on its turn that

\begin{equation}\label{eq:ideide}
\begin{array}{lll}
\frac{\partial \mathbf{g}^{(1)}( \mathbf{x}_i, \mathbf{x}_j)}{\partial  \mathbf{x}_i }   \bigg|_{(\mathbf{x}^s, \mathbf{x}^s) } +  \frac{\partial  \mathbf{g}^{(1)}( \mathbf{x}_i, \mathbf{x}_j) }{\partial  \mathbf{x}_j } \bigg|_{(\mathbf{x}^s, \mathbf{x}^s) } & = & 0,
\\
\frac{\partial \mathbf{g}^{(2)}( \mathbf{x}_i, \mathbf{x}_j, \mathbf{x}_k)}{\partial  \mathbf{x}_i }   \bigg|_{(\mathbf{x}^s,\mathbf{x}^s, \mathbf{x}^s) } + \frac{\partial \mathbf{g}^{(2)}(  \mathbf{x}_i, \mathbf{x}_j, \mathbf{x}_k)}{\partial  \mathbf{x}_j }   \bigg|_{(\mathbf{x}^s,\mathbf{x}^s, \mathbf{x}^s) } + & & \\+  \frac{\partial  \mathbf{g}^{(2)}(  \mathbf{x}_i, \mathbf{x}_j, \mathbf{x}_k) }{\partial  \mathbf{x}_k } \bigg|_{(\mathbf{x}^s,\mathbf{x}^s, \mathbf{x}^s) } & = & 0.
\end{array}
\end{equation}

Then, one can factor out the terms $\frac{\partial \mathbf{g}^{(1)}( \mathbf{x}_i, \mathbf{x}_j)}{\partial  \mathbf{x}_i }   \bigg|_{(\mathbf{x}^s, \mathbf{x}^s) } \delta \mathbf{x}_i$ and $\frac{\partial \mathbf{g}^{(2)}( \mathbf{x}_i, \mathbf{x}_j, \mathbf{x}_k)}{\partial  \mathbf{x}_i }   \bigg|_{(\mathbf{x}^s,\mathbf{x}^s, \mathbf{x}^s) } \delta \mathbf{x}_i$ in the summations (both of them, indeed, do not depend on the indices of the summations).
Furthermore, one has that $\sum_{j=1}^{N} a^{(1)}_{ij}=k_i^{(1)}$ and $\sum_{j=1}^N \sum_{k=1}^N a_{ijk}^{(2)}=2 k_i^{(2)}$.
Plugging back the resulting terms inside the summations, and using Eq. (\ref{eq:ideide}), one eventually obtains

\begin{equation}\label{trasformata}
\begin{array}{lll}
\dot{ \delta \mathbf{x}}_i & = &   J  \mathbf{f}(\mathbf{x}^s)   \delta \mathbf{x}_i - \sigma_{1} \sum\limits_{j=1}^{N} \mathcal{L}^{(1)}_{ij}  \: J \mathbf{g}^{(1)}( \mathbf{x}^s,\mathbf{x}^s) \delta \mathbf{x}_j \\
& & - \sigma_{2}  \sum\limits_{j=1}^N \sum_{k=1}^N \tau_{ijk} \:  \biggl [ J_1  \mathbf{g}^{(2)} (\mathbf{x}^s,\mathbf{x}^s, \mathbf{x}^s) \delta \mathbf{x}_j  \\
& & + J_2  \mathbf{g}^{(2)} (\mathbf{x}^s, \mathbf{x}^s, \mathbf{x}^s) \delta \mathbf{x}_k \biggr ] ,
\end{array}
\end{equation}

\noindent where we introduced a tensor $\mathrm{T}$ whose elements are $\tau_{ijk}=2  k_i^{(2)} \delta_{ijk} - a_{ijk}^{(2)}$ for $i,j,k=1, \ldots, N$, and simplified the notation as

\begin{equation}
\begin{array}{l}
J \mathbf{g}^{(1)} (\mathbf{x}^s,\mathbf{x}^s) =  \frac{\partial \mathbf{g}^{(1)}(  \mathbf{x}_i, \mathbf{x}_j)}{\partial  \mathbf{x}_j }   \bigg|_{(\mathbf{x}^s,\mathbf{x}^s)} ,
\\
J_1  \mathbf{g}^{(2)} (\mathbf{x}^s, \mathbf{x}^s, \mathbf{x}^s)=  \frac{\partial \mathbf{g}^{(2)}(\mathbf{x}_i, \mathbf{x}_j, \mathbf{x}_k)}{\partial  \mathbf{x}_j }   \bigg|_{(\mathbf{x}^s, \mathbf{x}^s, \mathbf{x}^s) }, \\
J_2  \mathbf{g}^{(2)} (\mathbf{x}^s, \mathbf{x}^s, \mathbf{x}^s)=  \frac{\partial \mathbf{g}^{(2)}(\mathbf{x}_i, \mathbf{x}_j, \mathbf{x}_k)}{\partial  \mathbf{x}_k }   \bigg|_{(\mathbf{x}^s, \mathbf{x}^s, \mathbf{x}^s) }.
\end{array}
\end{equation}

Already at this stage, it is fundamental to remark that our approach even extends the validity of the classical Master Stability Function theory [the case $\sigma_2=0$ in Eq. (\ref{trasformata})], in that {\it we do not require} a diffusive functional form for the interplay among the network nodes, and therefore we are actually encompassing a much broader class of coupling functions. For instance, our approach allows the formal treatment of the Kuramoto model \cite{acebron2005kuramoto}, where $m=1$, each network unit $i$ is identified by the instantaneous phase $\theta_i$ of an oscillator, and the coupling between nodes $i$ and $j$ is given by the function $\sin{(\theta_j-\theta_i)}$, which is not diffusive.

Let us now make our second, conceptual, step, which will allow us to greatly simplify the last term on the right hand side of Eq.(\ref{trasformata}). Such a term refers to three-body interactions, and we now show how to map it into a single summation involving the generalized Laplacian matrix.
This is done by remarking that the two Jacobian matrices $J_1  \mathbf{g}^{(2)} (\mathbf{x}^s, \mathbf{x}^s, \mathbf{x}^s)$ and $J_2  \mathbf{g}^{(2)} (\mathbf{x}^s, \mathbf{x}^s, \mathbf{x}^s)$ are both independent on $k$ and $j$. Accordingly,  Eq.(\ref{trasformata}) becomes

\begin{equation}
\label{eq:perturbations2a}
\begin{array}{lll}
\dot{ \delta \mathbf{x}}_i & = &  J  \mathbf{f}(\mathbf{x}^s)   \delta \mathbf{x}_i - \sigma_{1} \sum\limits_{j=1}^{N} \mathcal{L}^{(1)}_{ij}  \: J \mathbf{g}^{(1)}( \mathbf{x}^s, \mathbf{x}^s) \delta \mathbf{x}_j \\
& & - \sigma_{2} \biggl [  \sum\limits_{j=1}^N   J_1  \mathbf{g}^{(2)} (\mathbf{x}^s,\mathbf{x}^s, \mathbf{x}^s) \delta \mathbf{x}_j   \sum\limits_{k=1}^N \tau_{ijk} \\
& & +  \sum\limits_{k=1}^N  J_2  \mathbf{g}^{(2)} (\mathbf{x}^s,\mathbf{x}^s, \mathbf{x}^s) \delta \mathbf{x}_k  \sum\limits_{j=1}^N \tau_{ijk}  \biggr ].
  \end{array}
\end{equation}

Then, using the symmetric property of $ \mathrm{T}$, namely $
\sum_k \tau_{ijk}= \sum_k \tau_{ikj} $, we have

\begin{equation}
\label{eq:perturbations2}
\begin{array}{lll}
\dot{ \delta \mathbf{x}}_i & = &  J  \mathbf{f}(\mathbf{x}^s)   \delta \mathbf{x}_i - \sigma_{1} \sum\limits_{j=1}^{N} \mathcal{L}^{(1)}_{ij}  \: J \mathbf{g}^{(1)}( \mathbf{x}^s, \mathbf{x}^s) \delta \mathbf{x}_j \\
 & & - \sigma_{2} \biggl [  \sum\limits_{j=1}^N   J_1  \mathbf{g}^{(2)} (\mathbf{x}^s,\mathbf{x}^s, \mathbf{x}^s) \delta \mathbf{x}_j   \mathcal{L}^{(2)}_{ij}  \\
& & +  \sum\limits_{k=1}^N  J_2  \mathbf{g}^{(2)} (\mathbf{x}^s,\mathbf{x}^s, \mathbf{x}^s) \delta \mathbf{x}_k  \mathcal{L}^{(2)}_{ik}  \biggl ]  \\
 & = &  J  \mathbf{f}(\mathbf{x}^s)   \delta \mathbf{x}_i - \sigma_{1} \sum\limits_{j=1}^{N} \mathcal{L}^{(1)}_{ij}  \: J \mathbf{g}^{(1)}( \mathbf{x}^s, \mathbf{x}^s) \delta \mathbf{x}_j  \\
 & & - \sigma_{2} \sum\limits_{j=1}^N  \mathcal{L}^{(2)}_{ij}    \biggl [
  J_1  \mathbf{g}^{(2)} (\mathbf{x}^s,\mathbf{x}^s, \mathbf{x}^s) \\
  & & +  J_2  \mathbf{g}^{(2)} (\mathbf{x}^s,\mathbf{x}^s, \mathbf{x}^s)  \biggr ]  \delta  \mathbf{x}_j.
  \end{array}
\end{equation}

Let us now rewrite Eq. \eqref{eq:perturbations2} in  block form by introducing the stack vector $\delta \mathbf{x}=[\delta \mathbf{x}_1^T,\delta \mathbf{x}_2^T,\ldots,\delta \mathbf{x}_N^T]^T$ and denoting by $J \mathbf{F}= J  \mathbf{f}(\mathbf{x}^s)$, $ \mathrm{JG}^{(1)} =  J \mathbf{g}^{(1)}( \mathbf{x}^s, \mathbf{x}^s)$ and $ \mathrm{JG}^{(2)} =  J_1  \mathbf{g}^{(2)} (\mathbf{x}^s,\mathbf{x}^s, \mathbf{x}^s) +  J_2  \mathbf{g}^{(2)} (\mathbf{x}^s,\mathbf{x}^s, \mathbf{x}^s) $. One obtains

\begin{equation}\label{eq:blockEquation}
\dot{ \delta \mathbf{x}} =  \left [
\mathrm{I}_N \otimes   \mathrm{JF}  -
\sigma_1   \mathcal{L}^{(1)}  \otimes  \mathrm{JG}^{(1)} -
 \sigma_2  \mathcal{L}^{(2)} \otimes \mathrm{JG}^{(2)}  \right ]  \delta  \mathbf{x}.
\end{equation}

\begin{figure}
\subfigure[]{\includegraphics[width=0.23\textwidth]{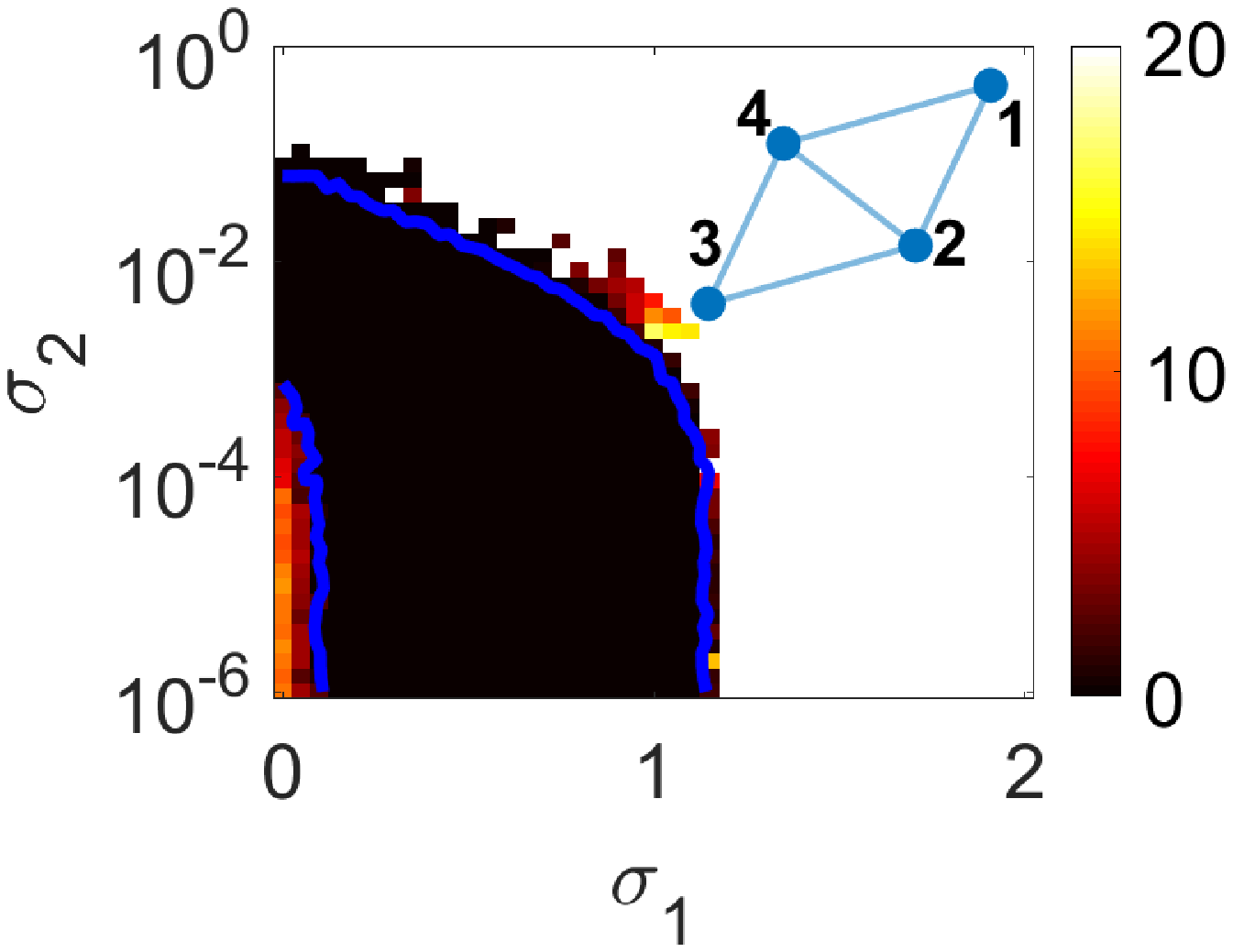}}
\subfigure[]{\includegraphics[width=0.23\textwidth]{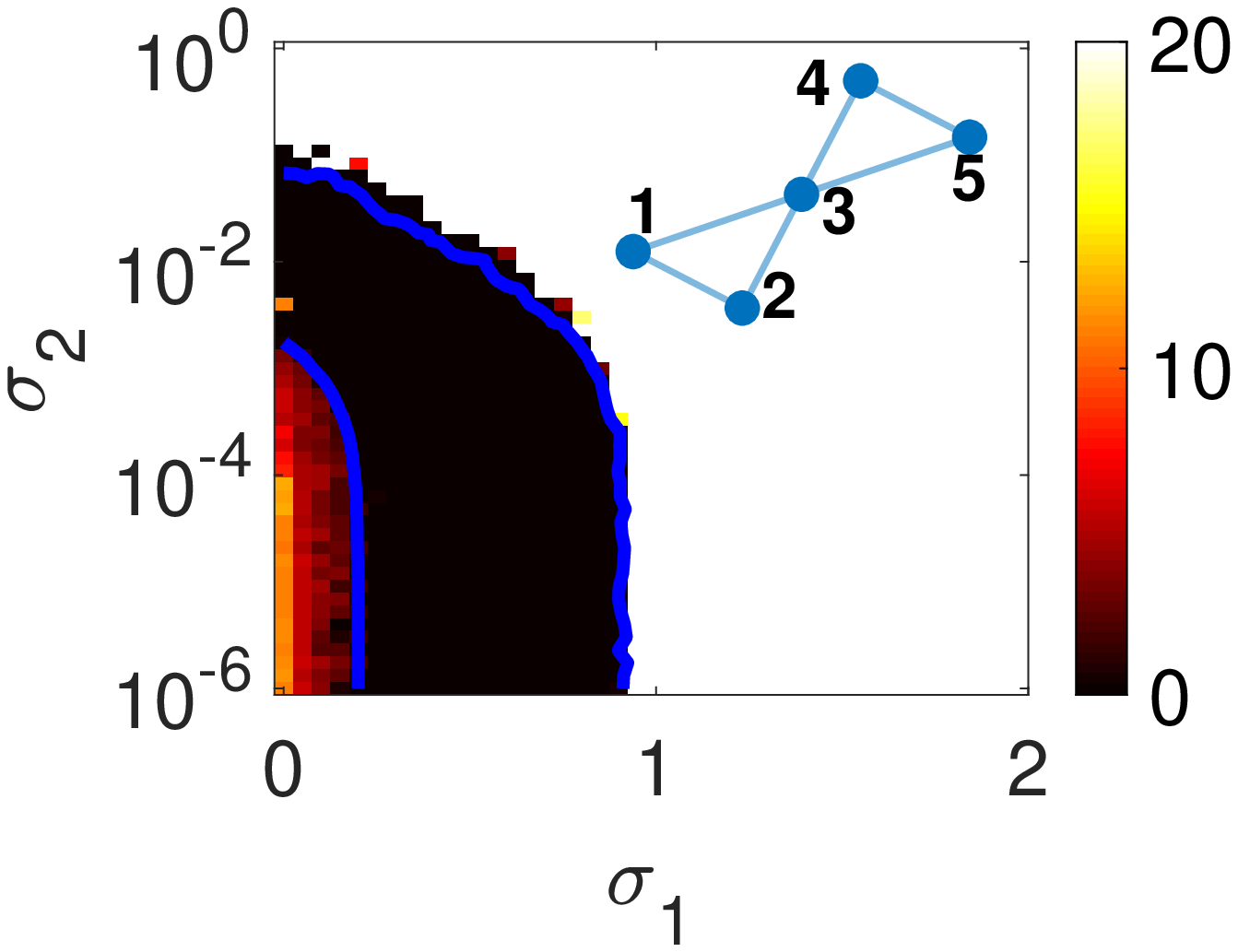}}
\subfigure[]{\includegraphics[width=0.23\textwidth]{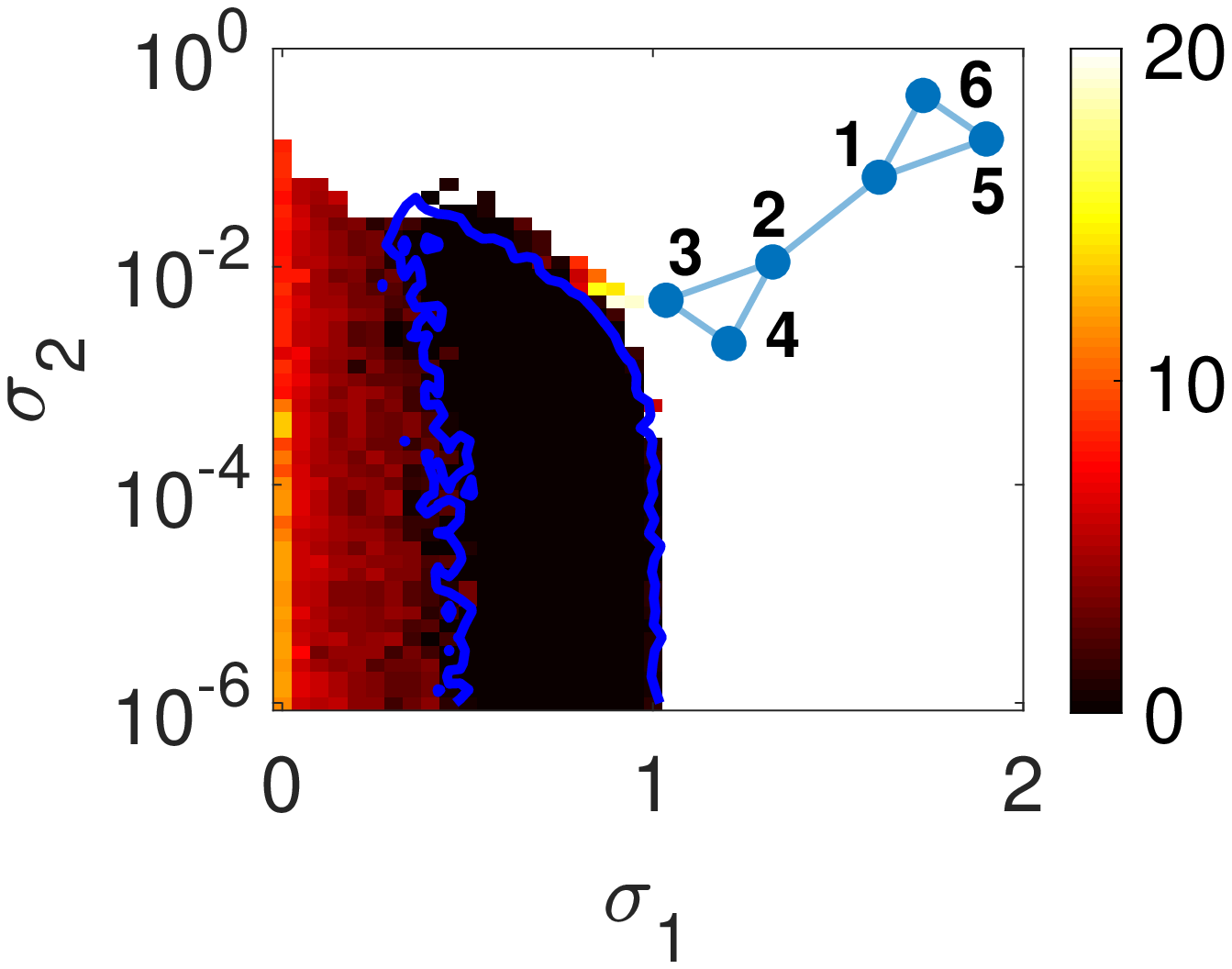}}
\subfigure[]{\includegraphics[width=0.23\textwidth]{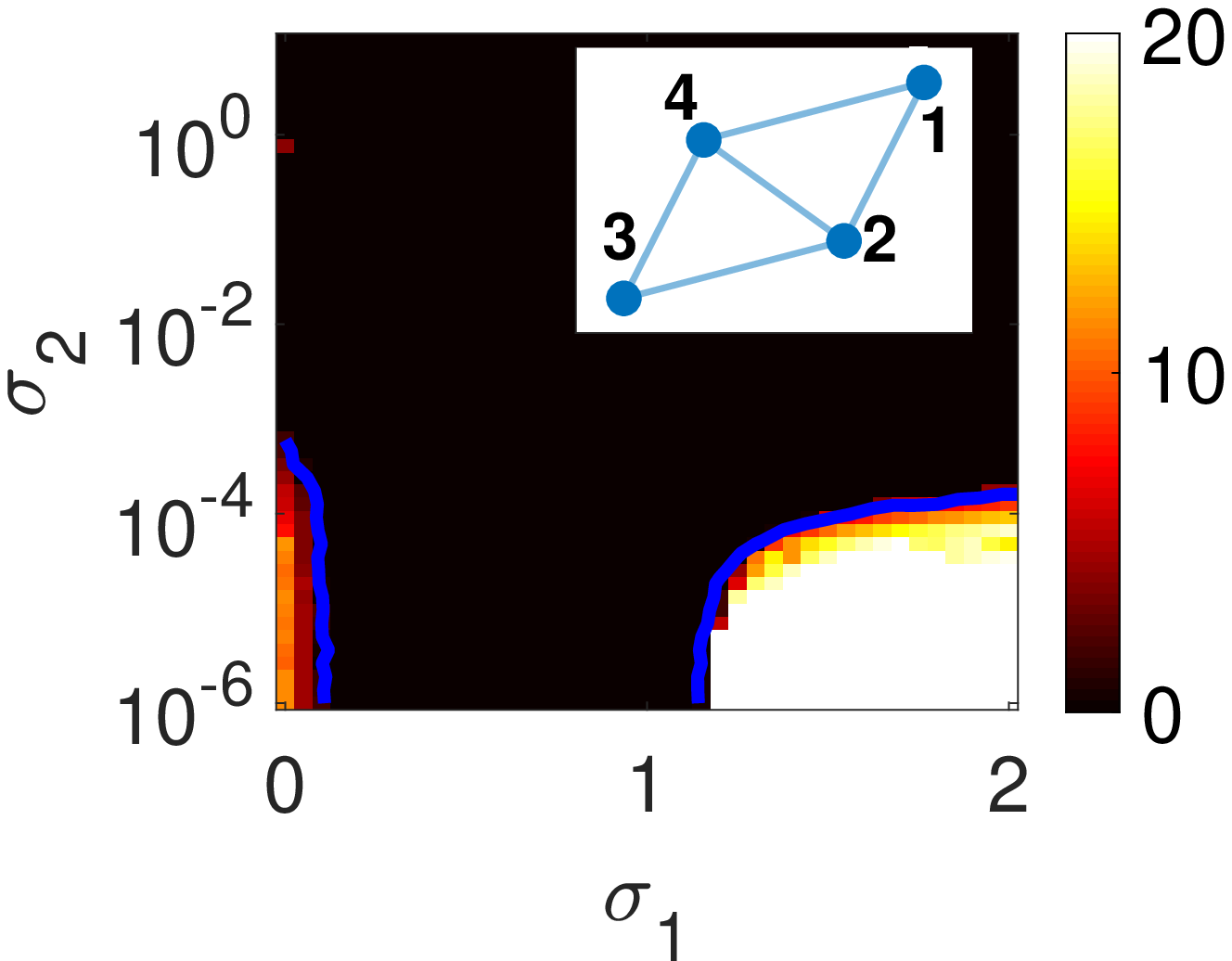}}
\subfigure[]{\includegraphics[width=0.23\textwidth]{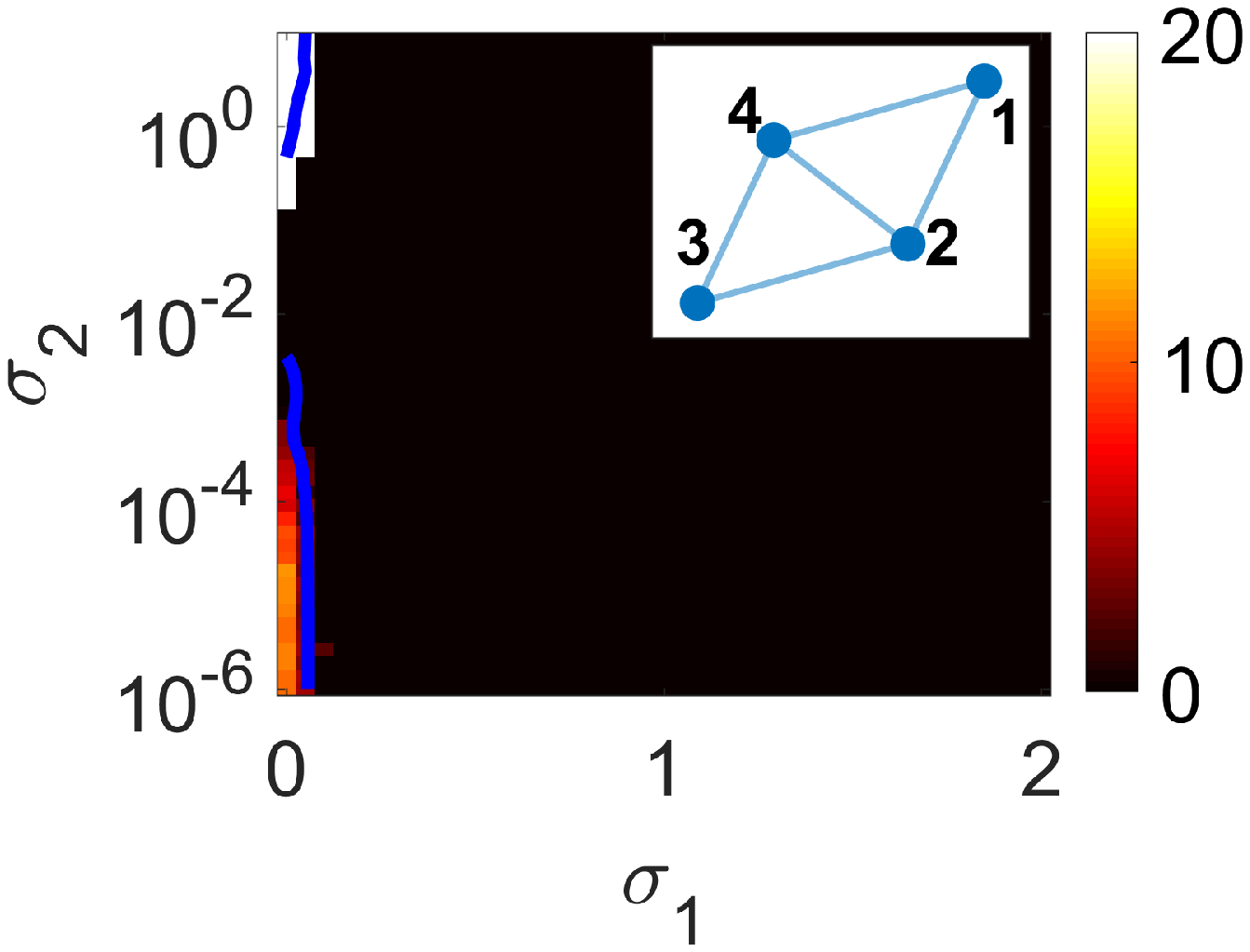}}
\caption{\label{fig:casogenerale} Contour plots of the time averaged (over an observation time $T=500$) synchronization error $E$ (see Eq. (\ref{syncherror}) for definition and the vertical bars of each panel for the color code) in the plane ($\sigma_1$, $\sigma_2$) for some examples of simplicial complexes (whose sketches are reported in the top left of each panel).  Simulations refer to coupled
R\"ossler oscillators ($\mathbf{x}=(x,y,z)^T$ and $\mathbf{f}=(-y -z, x + a y,  b + z (x -c))^T$) with parameters fixed in the chaotic regime ($a=b=0.2$, $c=9$). In panels (a-d) $\mathbf{g}^{(1)}(\mathbf{x}_i,\mathbf{x}_j)=[x_j-x_i,0,0]^T$, while in panel (e)  $\mathbf{g}^{(1)}(\mathbf{x}_i,\mathbf{x}_j)=[0,y_j-y_i,0]^T$.
As for the other coupling function, one has  $\mathbf{g}^{(2)}(\mathbf{x}_i,\mathbf{x}_j,\mathbf{x}_k)= [0,y_j^2y_k-y_i^3,0]^T$ in panel (d) and $\mathbf{g}^{(2)}(\mathbf{x}_i,\mathbf{x}_j,\mathbf{x}_k)=[x_j^2x_k-x_i^3,0,0]^T$ in all other panels. The blue continuous lines are the theoretical predictions of the synchronization thresholds obtained from Eq. \eqref{eq:MSFcasogenerale}. The three panels of the top row are examples of class III problems, whereas the two panels in the bottom row are examples of class II problems.}
\end{figure}

The third, and final, conceptual step is to remark that all generalized Laplacians $\mathcal{L}^{(d)}$ are symmetric real-valued zero-row-sum matrices. Therefore:
{\it (i)} they are all diagonalizable; {\it (ii)} for each one of them the set of eigenvalues is made of real non-negative numbers, and the corresponding set of eigenvectors constitutes a orthonormal basis of $\mathbb{R}^N$; {\it (iii)} they all share, as the smallest of their eigenvalues, $\lambda_1 \equiv 0$, whose associated eigenvector $\frac{1}{\sqrt{N}} \ (1,1,1,...,1)^T$ is aligned along the synchronization manifold; {\it (iv)} as in general they do not commute, the sets of eigenvectors corresponding to all others of their eigenvalues are different from one another, and yet any perturbation to the synchronization manifold (which, by definition, lies in the tangent space) can be expanded as linear combination of one whatever of such eigenvector sets (the relevant consequence is that one can arbitrarily select any of the generalized Laplacians as the reference for the choice of the basis of the transverse space, and all other eigenvector sets will map to such a basis by means of unitary matrix transformations).

We are then fully entitled to take, as reference basis, the one constituted by the eigenvectors of the classic Laplacian $\mathcal{L}^{(1)}$  ($\mathrm{V}=[\mathbf{v}_1,\mathbf{v}_2,\ldots,\mathbf{v}_N]$), and consider new variables $\mathbf{\delta\eta}=(\mathrm{V}^{-1}\otimes \mathrm{I}_N)\mathbf{\delta x}$. We get

\begin{equation}\label{eq:blockEquation2}
\nonumber
\begin{array}{lll}
\dot{ \mathbf{\delta\eta}} & = &  (\mathrm{V}^{-1}\otimes \mathrm{I}_N)\biggl [
\mathrm{I}_N \otimes   \mathrm{JF}  -
\sigma_1   \mathcal{L}^{(1)} \otimes  \mathrm{JG}^{(1)} \\
&& -
 \sigma_2  \mathcal{L}^{(2)} \otimes \mathrm{JG}^{(2)}  \biggr ] (\mathrm{V}\otimes \mathrm{I}_N)   \mathbf{\delta\eta}.
 \end{array}
\end{equation}

\noindent Furthermore, taking into account that $\mathrm{V}^{-1}\mathcal{L}^{(1)}\mathrm{V}=diag(\lambda_1,\lambda_2,\ldots,\lambda_N)=\mathrm{\Lambda}^{(1)}$, where $0=\lambda_1<\lambda_2\leq \ldots \lambda_N$ are the eigenvalues of $\mathcal{L}^{(1)}$, and indicating with $\tilde{\mathcal{L}}^{(2)}=\mathrm{V}^{-1}\mathcal{L}^{(2)}\mathrm{V}$ the transformed generalized Laplacian of order 2, one obtains that

\begin{equation}\label{eq:blockEquation3}
\dot{ \delta \mathbf{\eta}} =  \left [
\mathrm{I}_N \otimes   \mathrm{JF}  -
\sigma_1   \mathrm{\Lambda}^{(1)} \otimes  \mathrm{JG}^{(1)} -
 \sigma_2  \tilde{\mathcal{L}}^{(2)} \otimes \mathrm{JG}^{(2)}  \right ]  \delta  \mathbf{\eta}.
\end{equation}

As $\mathcal{L}^{(2)}$ is zero-row sum (i.e. $\mathcal{L}^{(2)}\mathbf{v}_1=0$), Eqs. \eqref{eq:blockEquation3} may be rewritten as

\begin{equation}
\label{eq:MSFcasogenerale}
\begin{cases}
\dot{\mathbf{\eta}}_1 & = \mathrm{JF}\mathbf{\eta}_1\\
\dot{  \mathbf{\eta}_i} & =  (\mathrm{JF}-\sigma_1 \lambda_i  \mathrm{JG}^{(1)} ) \mathbf{\eta}_i - \sigma_2 \sum\limits_{j=2}^N \tilde{\mathcal{L}}^{(2)}_{ij}\mathrm{JG}^{(2)}    \mathbf{\eta}_j,
\end{cases}
\end{equation}

\noindent that is, the dynamics of the linearized system is decoupled into two parts: the dynamics of $\mathbf{\eta}_1$ accounting for the motion along the synchronous manifold, and that of all other variables $\mathbf{\eta}_i$ (with $i=2,\ldots,N$, representing the different modes transverse to the synchronization manifold) which are coupled each other by means of the coefficients $\tilde{\mathcal{L}}^{(2)}_{ij}$ (all of them being known quantities) given by transforming $\mathcal{L}^{(2)}$ with the matrix that diagonalizes $\mathcal{L}^{(1)}$. The problem of stability is then reduced to: {\it (i)} simulating {\it a single, uncoupled,} nonlinear system;  {\it (ii)} using the obtained trajectory to feed up the elements of the Jacobians $\mathrm{JG}^{(1)}$ and $\mathrm{JG}^{(2)}$; {\it (iii)} simulating the dynamics of a system of $N-1$ coupled linear equations, and tracking the behavior of the norm $\sqrt{\sum_{i=2}^N \sum_{j=1}^m (\eta_i^{(j)})^2}$
for the calculation of the maximum Lyapunov exponent
(being $\mathbf{\eta}_{i} \equiv (\eta_i^{(1)}, \eta_i^{(2)}, ...,\eta_i^{(m)})$).

Stability of the synchronous solution requires, as a necessary condition, that the maximum among the Lyapunov exponents associated to all transverse modes be negative. Therefore, this quantity provides a generalized Master Stability Function, $\Lambda_{max}$, which, given the node dynamics and the coupling functions, is in general function of the topology of the two body interactions, the topology of the three body interactions, and the two coupling strengths $\sigma_1$ and $\sigma_2$, i.e., $\Lambda_{max}$=$\Lambda_{max}(\sigma_1,\sigma_2,\mathcal{L}^{(1)},\mathcal{L}^{(2)})$.

In analogy with the classification of systems made for synchronization of complex networks (Chapter 5 in Ref. \cite{boccaletti2006complex}), one immediately realizes that, once specified the dynamical system taking place in each node (i.e. the function $\mathbf{f}$), the various coupling functions $\mathbf{g}^{(1,2)}$, and the structure of the simplicial complex (i.e. $\mathcal{L}^{(1)}$ and $\mathcal{L}^{(2)}$), all possible cases can be divided in  three classes: {\it (i)} class I problems, where $\Lambda_{max}$ is positive in all the half plane $(\sigma_1 \geq 0,\sigma_2 \geq 0)$, and therefore synchronization is never stable; {\it (ii)} class II problems, for which $\Lambda_{max}$ is negative within a unbounded area of the half plane $(\sigma_1 \geq 0,\sigma_2 \geq 0)$; and {\it (iii)} class III problems, for which the area of the half plane $(\sigma_1 \geq 0,\sigma_2 \geq 0)$ in which $\Lambda_{max}$ is negative is instead bounded, and therefore additional instabilities of the synchronous motion may occur at larger values of the coupling strengths. While class I problems are trivial (in that synchronization is never observed), examples of class II and class III problems are shown in
Fig. \ref{fig:casogenerale} for simplicial complexes of R\"ossler oscillators \cite{rossler1976equation}, and one easily sees that the predictions made by solving Eqs. (\ref{eq:MSFcasogenerale}) are indeed fully confirmed by the simulations of the
original system equations (\ref{eq:msf}).

Far from being limited to the case of $D=2$, our approach can be extended straightforwardly to simplicial complexes of any order $D$. Each term on the right hand side of Eq. (\ref{eq:general}) can, indeed, be manipulated following exactly the same three conceptual steps described so far.
Calling therefore $ \mathrm{JG}^{(d)} =  J_1 \mathbf{g}^{(d)} (\mathbf{x}^s,..., \mathbf{x}^s) +  J_2  \mathbf{g}^{(d)} (\mathbf{x}^s,..., \mathbf{x}^s) + ...+ J_d  \mathbf{g}^{(d)} (\mathbf{x}^s,..., \mathbf{x}^s)$, Eq. (\ref{eq:blockEquation}) becomes

\begin{equation}\label{eq:blockEquationgeneral}
\begin{array}{lll}
\dot{ \delta \mathbf{x}} & = &  \biggl [
\mathrm{I}_N \otimes   \mathrm{JF}  -
\sigma_1   \mathcal{L}^{(1)}  \otimes  \mathrm{JG}^{(1)} -
 \sigma_2  \mathcal{L}^{(2)} \otimes \mathrm{JG}^{(2)}   - ... \\
 & & - \sigma_D \mathcal{L}^{(D)} \otimes \mathrm{JG}^{(D)} \biggr ]  \delta  \mathbf{x}.
 \end{array}
\end{equation}

Once again, one is entitled to select the eigenvector set which diagonalizes $\mathcal{L}^{(1)}$, and to introduce the new variables $\mathbf{\delta\eta}=(\mathrm{V}^{-1}\otimes \mathrm{I}_n)\mathbf{\delta x}$.
Following the very same steps which led us to write Eqs. (\ref{eq:MSFcasogenerale}), one then obtains

\begin{equation}
\label{eq:MSFcasogenerale2}
\begin{array}{lll}
\dot{\mathbf{\eta}}_1 & = & \mathrm{JF}\mathbf{\eta}_1,\\
\dot{\mathbf{\eta}_i} & = & (\mathrm{JF}-\sigma_1 \lambda_i  \mathrm{JG}^{(1)} ) \mathbf{\eta}_i - \sigma_2 \sum\limits_{j=2}^N \tilde{\mathcal{L}}^{(2)}_{ij}\mathrm{JG}^{(2)} \mathbf{\eta}_j   - \ldots \\
& & - \sigma_D \sum\limits_{j=2}^N \tilde{\mathcal{L}}^{(D)}_{ij}\mathrm{JG}^{(D)} \mathbf{\eta}_j , \end{array}
\end{equation}
where the coefficients $\tilde{\mathcal{L}}^{(d)}_{ij}$ result from transforming $\mathcal{L}^{(d)}$ with the matrix that diagonalizes $\mathcal{L}^{(1)}$.
As a result, one has conceptually the same reduction of the problem to  {\it a single, uncoupled,} nonlinear system, plus a system of $N-1$ coupled linear equations, from which the maximum Lyapunov exponent $\Lambda_{max}$=$\Lambda_{max}(\sigma_1,\sigma_2,..., \sigma_D,\mathcal{L}^{(1)},\mathcal{L}^{(2)}, ..., \mathcal{L}^{(D)})$ can be extracted and monitored (for each simplicial complex) in the $D-$dimensional hyper-space of the coupling strength parameters.

\section{Special cases}
\label{sec:specialcases}

Going back to the case of $D=2$ (once again only for the sake of illustration, as all exemplifications we will make are straightforwardly extendable to any order $D$), the problem can be greatly simplified in a few special cases in which either the topology of the connectivity structure, or the coupling functions, allow for a further reduction of complexity.

\subsection{All-to-all coupling}

The first case is an all-to-all coupling, for which every two and three-body interaction is active. In this case, the classical Laplacian matrix is
\begin{equation*}
 \mathcal{L}_{ij}^{(1)} =
\begin{cases}
-1  & \quad \text{ for } i \neq j \\
N-1  & \quad   \text{ for } i = j.
\end{cases}
\end{equation*}
Then, it is easy to rewrite $\mathcal{L}^{(2)}$, because the off diagonal terms $\mathcal{L}^{(2)}_{ij}$ ($i \neq j$) represent the number of triangles formed by the link $(i,j)$ which, in the present case, is simply equal to $N-2$. Second, we consider the terms of the main diagonal $\mathcal{L}^{(2)}_{ii}$, the number of triangles having the node $i$ as a vertex, which is
\begin{equation*}
 k^{(2)}_{i} = \binom{N-1}{2} = \frac{(N-1)(N-2)}{2}.
\end{equation*}

Consequently, one has that
\begin{equation}\label{eq:a2a_laplacians}
\nonumber
 \mathcal{L}^{(2)} = (N-2) \; \mathcal{L}^{(1)}.
\end{equation}

With the current notation, one has therefore
\begin{eqnarray*}
\dot{ \delta \mathbf{x}}_i  =   J  \mathbf{F}   \delta \mathbf{x}_i - \sum_{j=1}^{N} \mathcal{L}_{ij}^{(1)} \: \left[\sigma_{1}\: J \mathbf{G}^{(1)} + \sigma_{2}\:(N-2)\:
  J  \mathbf{G}^{(2)} \right]     \delta  \mathbf{x}_j.
\end{eqnarray*}
By expanding the perturbation vector $\delta \mathbf{x}$ on the othornormal basis formed by the eigenvectors of the classical Laplacian matrix $\mathcal{L}^{(1)}$, and after noticing that in the all-to-all configuration $\lambda_{2} = \dots \lambda_{N} = N$, for each $\mathbf{\eta}_{i}$ (with $i \in \left\{2,\dots,N\right\}$) one has
\begin{equation}
\label{eq:blocchiall2all}
\dot{\mathbf{\eta}}_i=[\mathrm{JF}-\sigma_1 N \: JG^{(1)} -\sigma_2 N(N-2)\:JG^{(2)}]\mathbf{\eta}_i .
\end{equation}

In other words, in the all-to-all case, the variables $\mathbf{\eta}_{i}$ come out to be all uncoupled to each other, so that the MSF uniquely depends on $\sigma_1$, $\sigma_2$ and $N$, i.e., $\Lambda_{max}=\Lambda_{max}(\sigma_1,\sigma_2,N)$.

\subsection{Generalized diffusion interactions with natural coupling}

Another interesting case is that of generalized diffusion interactions with {\it natural} coupling functions.
This amounts to consider diffusive coupling functions, given by
\begin{equation} \label{eq:oneless}
\begin{split}
\mathbf{g}^{(1)}(\mathbf{x}_i, \mathbf{x}_j) &= \mathbf{h}^{(1)}( \mathbf{x}_j) - \mathbf{h}^{(1)}(\mathbf{x}_i), \\
\mathbf{g}^{(2)}(\mathbf{x}_i, \mathbf{x}_j, \mathbf{x}_k) &= \mathbf{h}^{(2)}( \mathbf{x}_j, \mathbf{x}_k) - \mathbf{h}^{(2)}(\mathbf{x}_i, \mathbf{x}_i),
\end{split}
\end{equation}

\noindent where  $\mathbf{h}^{(1)}: \mathbb{R}^{m} \longrightarrow  \mathbb{R}^m$ and $\mathbf{h}^{(2)}: \mathbb{R}^{2m} \longrightarrow \mathbb{R}^m$.
In addition, a condition of natural coupling is considered:
\begin{equation}
\label{eq:defnaturalcoupling}
\mathbf{h}^{(2)}(\mathbf{x},\mathbf{x}) = \mathbf{h}^{(1)}(\mathbf{x}).
\end{equation}

Eq. (\ref{eq:defnaturalcoupling}) expresses, indeed, the fact that the coupling to node $i$ from two-body and three-body interactions is essentially similar, in that a three-body interaction where two nodes are on the same state is equivalent to a two-body interaction.
Here, the MSF assumes a particularly convenient form, as it can be written as a function of a single parameter.

The consequence of
Eq. (\ref{eq:defnaturalcoupling}) is that
$J_1  \mathbf{h}^{(2)} (\mathbf{x}^s, \mathbf{x}^s) +  J_2  \mathbf{h}^{(2)} (\mathbf{x}^s, \mathbf{x}^s) = J \mathbf{h}^{(1)}(\mathbf{x}^s)$. Therefore, one has
\begin{equation}
\begin{array}{lll}
\dot{ \delta \mathbf{x}}_i & = &  J  \mathbf{f}(\mathbf{x}^s)   \delta \mathbf{x}_i - \sigma_{1} \sum_{j=1}^{N} \mathcal{L}_{ij}^{(1)} \: J \mathbf{h}^{(1)}( \mathbf{x}^s) \delta \mathbf{x}_j \\ & & - \sigma_{2}  \sum_{j=1}^N \mathcal{L}_{ij}^{(2)} \: J \mathbf{h}^{(1)}( \mathbf{x}^s) \delta \mathbf{x}_{j}  \\
 & = &  J  \mathbf{f}(\mathbf{x}^s)   \delta \mathbf{x}_i - \sum_{j=1}^{N} \left [ \sigma_{1} \mathcal{L}_{ij}^{(1)} + \sigma_{2} \mathcal{L}_{ij}^{(2)} \right ] \: J \mathbf{h}^{(1)}( \mathbf{x}^s) \delta \mathbf{x}_j.
\end{array}
\end{equation}
Alternatively, one can consider the  zero-row-sum, symmetric, \textit{effective matrix} $\mathcal{M}$, given by
\begin{equation*}
\mathcal{M} = \mathcal{L}^{(1)} + r \mathcal{L}^{(2)},  \qquad r = \frac{\sigma_2}{\sigma_1}.
\end{equation*}
The eigenvalues of $\mathcal{M}$ depend on the ratio $r$ of the coupling coefficients, and one has that
\begin{equation}
\label{eq:MSFnatural}
\dot{ \delta \mathbf{x}}_i =  J  \mathbf{f}(\mathbf{x}^s)   \delta \mathbf{x}_i - \sigma_{1} \sum_{j=1}^{N} \mathcal{M}_{ij} \: J \mathbf{h}^{(1)}( \mathbf{x}^s) \delta \mathbf{x}_j.
\end{equation}

Eq. \eqref{eq:MSFnatural} allows to establish a formal full analogy between the case of a simplicial complex and that of a network with weights given by the coefficients of the effective matrix $\mathcal{M}$. In particular, a single-parameter MSF can be defined, starting from the following $m-dimensional$ linear parametric variational equation

\begin{equation}
\label{eq:MSFsimpliciale}
    \dot{\mathbf{\eta}} = \left[ J  \mathbf{f}(\mathbf{x}^s) - \alpha   J  \mathbf{h}^{(1)}(\mathbf{x}^s)  \right]  {\mathbf{\eta}}
\end{equation}

\noindent from which the maximum Lyapunov exponent is calculated: $\Lambda_{max}=\Lambda_{max}(\alpha)$ with  $\alpha=\lambda(\sigma_1\mathcal{L}^{(1)}+\sigma_2\mathcal{L}^{(2)})$ or $\alpha=\sigma_1\lambda(\mathcal{L}^{(1)}+r\mathcal{L}^{(2)})=\sigma_1\lambda(\mathcal{M})$.
The situation is therefore conceptually equivalent to that of synchronization in complex networks: given the dynamical system $\mathbf{f}$, the coupling functions $\mathbf{h}^{(1)}$ and $\mathbf{h}^{(2)}$, and the structure of connection of the simplicial complex (i.e. $ \mathcal{L}^{(1)}$ and $\mathcal{L}^{(2)}$) one can define three possible classes of problems: \begin{enumerate}[label=(\roman*)]
\item  class I problems, for which the curve $\Lambda_{max}=\Lambda_{max}(\alpha)$ does not intercept the abscissa and it is always positive. In this case synchronization is always forbidden, no matter which simplicial complex is used for connecting the dynamical systems;
\item class II problems, for which the curve $\Lambda_{max}=\Lambda_{max}(\alpha)$ intercepts the abscissa only once at $\alpha_c$, and for which, therefore, the synchronization threshold is given by the self consistent equation   $\sigma_1^{critical}= \alpha_c/ \lambda_2 [\mathcal{M}(\sigma_1^{critical},\sigma_2^{critical})]$, i.e. it scales with the inverse of the second smallest eigenvalue of the effective matrix;
\item  class III problems, for which the curve $\Lambda_{max}=\Lambda_{max}(\alpha)$ intercepts the abscissa twice at $\alpha_1$ and $\alpha_2 >\alpha_1$. In this case, synchronization can be observed only if the entire eigenvalue spectrum of the effective matrix is such that $\sigma_1 \lambda_2(\mathcal{M}) > \alpha_1$ and, at the same time, $\sigma_1 \lambda_N(\mathcal{M}) < \alpha_2$. In this case, the parameter $\frac{\lambda_2(\mathcal{M})}{\lambda_N(\mathcal{M}}$  can be considered as a proxy measure of {\it synchronizability} of the simplicial complex,  in that the closer is such a parameter to unity (the more compact is the spectrum of eigenvalue of $\mathcal{M}$) the larger can be the range of coupling strengths for which  the two above synchronization conditions can be satisfied.
\end{enumerate}

\section{Results}

We here discuss a series of results confirming the validity and wide applicability of the proposed approach. In particular, we will start with discussing the more general case dealt with in Section \ref{sec:generalcase}, and then we will focus on the special cases of Section \ref{sec:specialcases}. Moreover, we will focus on two paradigmatic examples of three-dimensional ($\mathbf{x}=(x,y,z)^T \in \mathbb{R}^3$) chaotic systems: the R\"ossler oscillator \cite{rossler1976equation}, and the Lorenz system \cite{strogatz2018nonlinear}. The R\"ossler oscillator is described by

\begin{equation}
\label{eq:rossler}
%\left \{
\begin{array}{l}
\dot{x} = -y -z,\\
\dot{y} = x + a y,  \\
\dot{z} = b + z (x -c),\\
\end{array}
%\right.
\end{equation}

\noindent while the equations for the Lorenz system are

\begin{equation}
\label{eq:lorenz}
%\left \{
\begin{array}{l}
\dot{x} = \sigma (y-x),\\
\dot{y} = x(\rho-z)-y ,\\
\dot{z} = xy-\beta z,\\
\end{array}
%\right.
\end{equation}

In both cases, the parameters are fixed so as the resulting dynamics is chaotic. Namely, for the R\"ossler oscillator we selected $a=b=0.2$, $c=9$, and for the Lorenz system  $\sigma=10$, $\rho=28$, and $\beta=8/3$.

Our discussion begins with going back to Fig. \ref{fig:casogenerale}, where we have considered a few elementary configurations of SCs, chosen in order to illustrate the classes of problems that one can deal with even when the structures involve only a small number of nodes. In particular, Fig. \ref{fig:casogenerale} reveals that synchronization in the general case crucially depends on the topology and the coupling functions: the same configuration can in fact feature different dynamics when diverse mechanisms regulate the coupling and, conversely, the same coupling functions may lead to different behaviors when the topology of interactions changes.

\begin{figure}[h]
\includegraphics[width=0.4\textwidth]{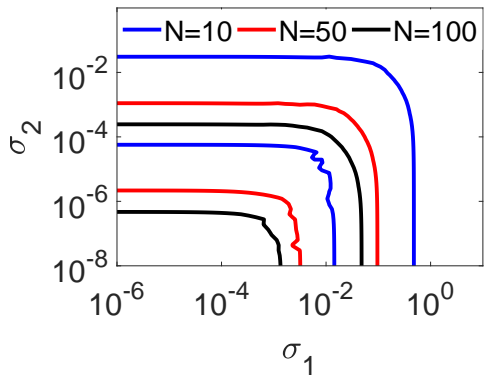}
\caption{\label{fig:all2all} Synchronization in a simplicial complex of R\"ossler oscillators with all-to-all coupling. Lower and upper boundary curves for the region where synchronization is stable, at different values of $N$. The color codes for the different curves is reported at the top of the panel.}
\end{figure}

As an example, let us consider the full dynamical equations of coupled R\"ossler oscillators, when the coupling functions are chosen as $\mathbf{g}^{(1)}(\mathbf{x}_i,\mathbf{x}_j)=[x_j-x_i,0,0]^T$ and $\mathbf{g}^{(2)}(\mathbf{x}_i,\mathbf{x}_j,\mathbf{x}_k)=[x_j^2x_k-x_i^3,0,0]^T$. They read

\begin{equation}
\label{eq:rosslernetscasogeneraleoldstyle}
%\left \{
\begin{array}{lll}
\dot{x}_{i} & = & -y_{i} -z_{i} +\sigma_1\sum\limits_{j=1}^N a_{ij}^{(1)}(x_j-x_i)\\
& & +\sigma_2\sum\limits_{j=1}^N\sum\limits_{k=1}^N a_{ij}^{(2)}(x_j^2x_k-x_i^3),\\
\dot{y}_{i} & = & x_{i} + a y_{i},  \\
\dot{z}_{i} & = & b + z_{i} (x_{i} -c),\\
\end{array}
%\right.
\end{equation}

\noindent or, equivalently

\begin{equation}
\label{eq:rosslernetscasogenerale}
%\left \{
\begin{array}{l}
\dot{x}_{i} = -y_{i} -z_{i} -\sigma_1\sum\limits_{j=1}^N\mathcal{L}_{ij}^{(1)}x_j-\sigma_2\sum\limits_{j=1}^N\sum\limits_{k=1}^N\tau_{ijk}x_j^2x_k,\\
\dot{y}_{i} = x_{i} + a y_{i},  \\
\dot{z}_{i} = b + z_{i} (x_{i} -c),\\
\end{array}
%\right.
\end{equation}

In each of the configurations considered, the state of the system is monitored by the average synchronization error defined as

\begin{equation}
\label{syncherror}
    E=\langle \left(\frac{1}{N(N-1)}\sum\limits_{i,j=1}^N \| \mathbf{x_j}-\mathbf{x_i}\|^2\right)^{\frac{1}{2}}\rangle_{T},
\end{equation}

\noindent where $T$ is a sufficiently large window of time where the synchronization error is averaged, after discarding the transient.

\begin{figure}
\includegraphics[width=0.5\textwidth]{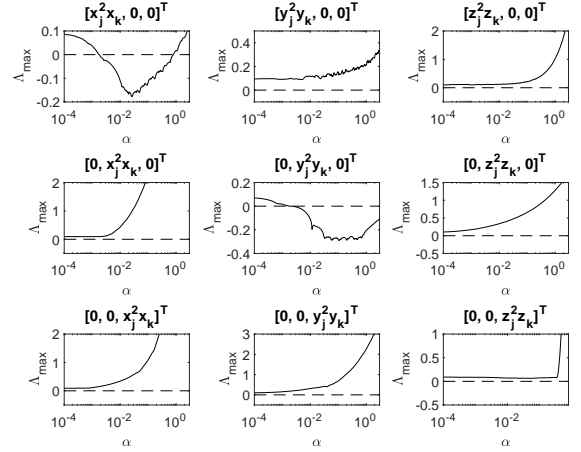}
\caption{\label{fig:MFS} Synchronization in simplicial complexes of R\"ossler oscillators, in the case of natural coupling. The Master Stability Function obtained for several coupling functions. On the top of each panel, the expression used for $\mathbf{h}^{(2)}$ is reported. The corresponding expression for $\mathbf{h}^{(1)}$ can be derived taking into account condition \eqref{eq:defnaturalcoupling}.}
\end{figure}

Fig.~\ref{fig:casogenerale} reports $E(\sigma_1,\sigma_2)$ for different SCs (shown as insets in the  panels) and coupling functions, along with the theoretical predictions provided by the MSF obtained from eq. \eqref{eq:MSFcasogenerale} (the blue, continuous, lines superimposed to the diagrams of the synchronization error). In all the cases, the numerical simulations are in very good agreement with the theoretical predictions for the synchronization thresholds.

Numerical integrations are performed by means of an Euler algorithm, with integration step $\delta t= 10^{-4}$, in a windows of time equal to $2T$ with $T=500s$. The MSF has been calculated from Eqs. (\ref{eq:MSFcasogenerale}) using the algorithm for the calculation of the maximum Lyapunov exponent reported in Ref. \cite{sprott2003chaos} (pp. 116-117) with the following parameters: integration step size $\delta t = 10^{-3}$, number of iterations per cycle $I = 10000$, number of cycles $C = 5$.

The results of Fig.~\ref{fig:casogenerale} suggest several interesting considerations. Indeed, in the cases reported in panels (a) and (b) of   Fig.~\ref{fig:casogenerale} synchronization may be achieved using either two-body or three-body interactions only (for very small $\sigma_1$ indeed there is a range of values of $\sigma_2$ leading to synchronization, and viceversa), while in the case of panel (c) synchronization is forbidden for very small values of $\sigma_1$. In the last case, in fact, the two triangles do not have a common edge as in Fig.~\ref{fig:casogenerale}(a), nor a common node as in Fig.~\ref{fig:casogenerale}(b), and therefore interactions through links becomes essential for synchronization. Finally, one notice that there are scenarios, as in panels (d) and (e), where the synchronization region is unbounded.  As already mentioned in Section \ref{sec:generalcase}, Fig.~\ref{fig:casogenerale} provides examples of two of the three possible classes of MSF, with class III behavior in Fig.~\ref{fig:casogenerale}(a)-(c), and class II in Fig.~\ref{fig:casogenerale}(d) (where synchronization exists in an unbounded region of the coupling coefficient regulating pairwise interactions, i.e., $\sigma_1$) and in Fig.~\ref{fig:casogenerale}(e) (where synchronization exists in an unbounded region of the coupling coefficient regulating three-body interactions, i.e., $\sigma_2$).

\begin{figure}
\includegraphics[width=0.5\textwidth]{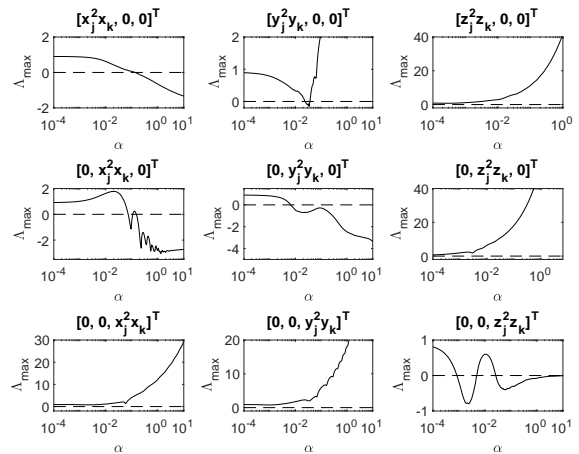}
\caption{\label{fig:MFSLorenz} Synchronization in simplicial complexes of Lorenz systems, in the case of natural coupling. The Master Stability Function is here calculated for several coupling functions. On the top of each panel, the expression used for $\mathbf{h}^{(2)}$ is reported. The corresponding expression for $\mathbf{h}^{(1)}$ can be derived taking into account condition \eqref{eq:defnaturalcoupling}.}
\end{figure}

Let us now move to discuss other results, which refer to the special cases of Section \ref{sec:specialcases}. We start with the all-to-all coupling case where, according to Eq. (\ref{eq:blocchiall2all}), one obtains a MSF that is function of $N$, $\sigma_1$ and $\sigma_2$. We then consider a simplicial complex of R\"ossler oscillators with all-to-all coupling, described by

\begin{equation}
\label{eq:rosslernetscasoAll2All}
%\left \{
\begin{array}{l}
\dot{x}_{i} = -y_{i} -z_{i} -\sigma_1\sum\limits_{j=1}^Nx_j-\sigma_2\sum\limits_{j=1}^N\sum\limits_{k=1}^Nx_j^2x_k,\\
\dot{y}_{i} = x_{i} + a y_{i} , \\
\dot{z}_{i} = b + z_{i} (x_{i} -c).\\
\end{array}
%\right.
\end{equation}

\noindent  The results are shown in Fig. \ref{fig:all2all} for three values of $N$ ($N=10$, $N=50$, and $N=100$): the synchronous manifold is stable in a bounded region of the semiplane $(\sigma_1>0,\sigma_2>0)$ delimited by blue ($N=10$), red ($N=50$) and black (N=$100$) lines. One immediately sees that such a stability region moves towards the origin when $N$ is increased. Hence, increasing $N$ reduces the lower and upper thresholds for achieving synchronization.

\begin{figure}
\subfigure[]{\includegraphics[width=0.23\textwidth]{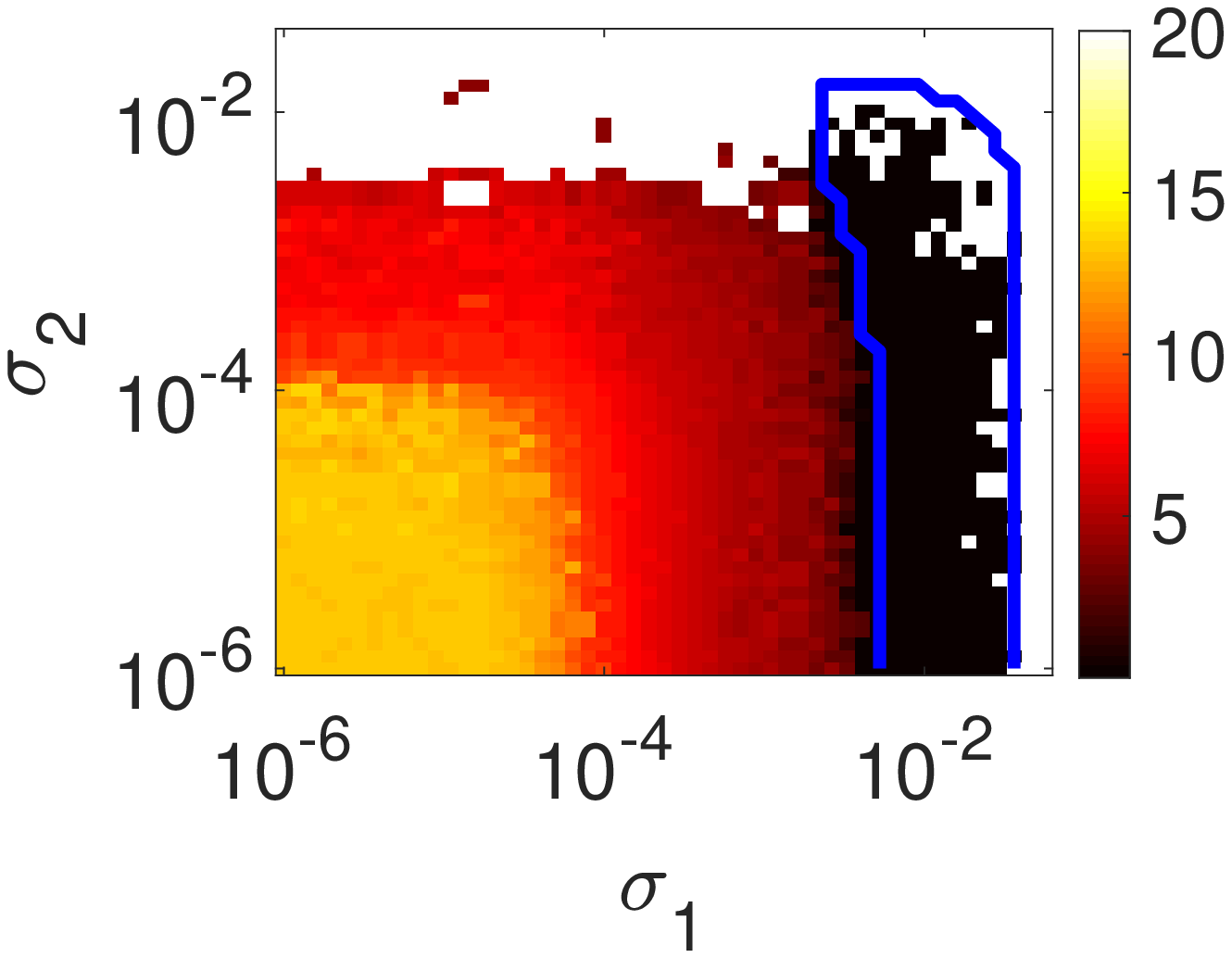}}
\subfigure[]{\includegraphics[width=0.23\textwidth]{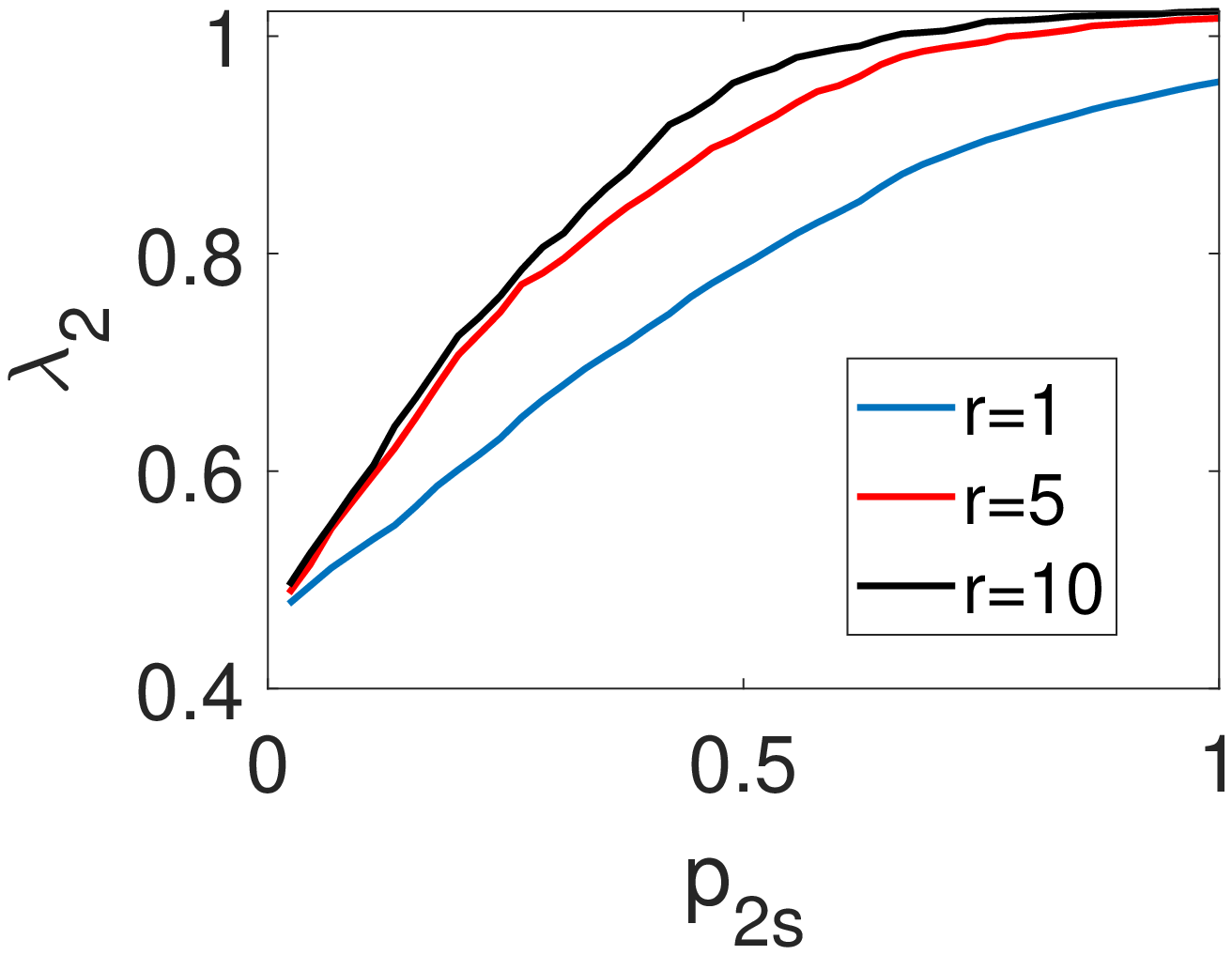}}
\subfigure[]{\includegraphics[width=0.23\textwidth]{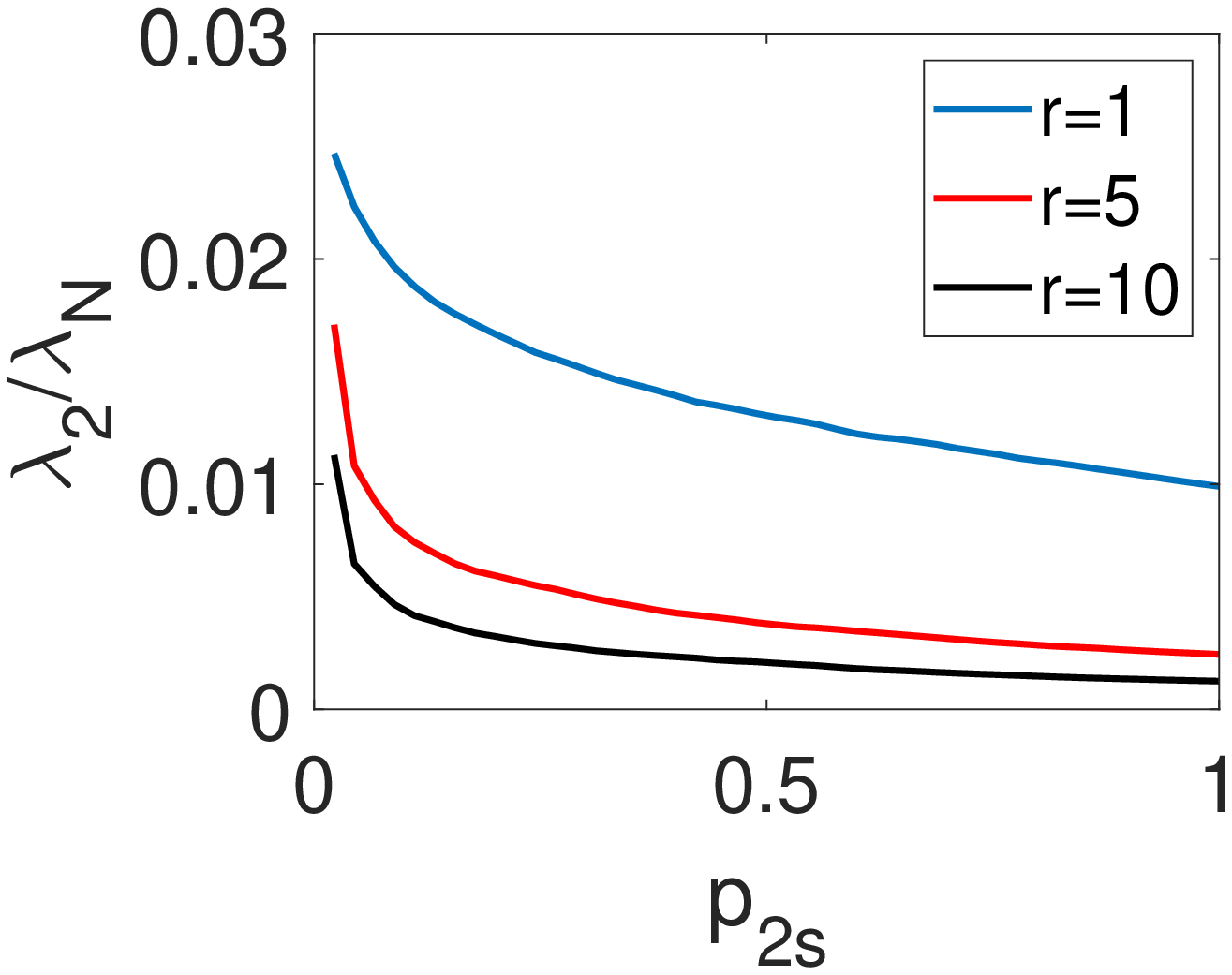}}\\
\caption{\label{fig:Zac} Synchronization in simplicial complexes extracted from the interactions characterizing the Zachary karate club network. (a) Synchronization error (color code reported in the bar at the right of the panel) vs. $\sigma_1$ and $\sigma_2$ for the simplicial complex obtained when all the triangles are considered as being 2-simplexes. The blue line delimits the area of stability of the synchronous solution predicted by the MSF. (b)  $\lambda_2$ vs. the percentage of 2-simplexes in the structure, $p_{2s}$ (see text for definition); (c) $\lambda_2/\lambda_N$ vs. $p_{2s}$. In panels (b) and (c) three different values of $r$ are considered, with the color code for the plotted curves being reported in the corresponding insets.}
\end{figure}

Finally, we consider the case of natural coupling. Here, as discussed in Sec. \ref{sec:specialcases}, the MSF is a function of a single parameter, i.e., $\Lambda_{max}=\Lambda_{max}(\alpha)$ with  $\alpha=\lambda(\sigma_1\mathcal{L}^{(1)}+\sigma_2\mathcal{L}^{(2)})$ or $\alpha=\sigma_1\lambda(\mathcal{L}^{(1)}+r\mathcal{L}^{(2)})=\sigma_1\lambda(\mathcal{M})$.

We calculated the MSF for the R\"ossler oscillator and the Lorenz system with several choices of the coupling functions:
$\mathbf{h}^{(1)}(\mathbf{x}_j)=[x_j^3,0,0]^T$ and $\mathbf{h}^{(2)}(\mathbf{x}_j,\mathbf{x}_k)=[x_j^2x_k,0,0]^T$;  $\mathbf{h}^{(1)}(\mathbf{x}_j)=[0,x_j^3,0]^T$ and $\mathbf{h}^{(2)}(\mathbf{x}_j,\mathbf{x}_k)=[0,x_j^2x_k,0]^T$; $\mathbf{h}^{(1)}(\mathbf{x}_j)=[0,0,x_j^3]^T$ and $\mathbf{h}^{(2)}(\mathbf{x}_j,\mathbf{x}_k)=[0,0,x_j^2x_k]^T$;  $\mathbf{h}^{(1)}(\mathbf{x}_j)=[y_j^3,0,0]^T$ and $\mathbf{h}^{(2)}(\mathbf{x}_j,\mathbf{x}_k)=[y_j^2y_k,0,0]^T$ ... $\mathbf{h}^{(1)}(\mathbf{x}_j)=[0,0,z_j^3]^T$ and $\mathbf{h}^{(2)}(\mathbf{x}_j,\mathbf{x}_k)=[0,0,z_j^2z_k]^T$. For the calculation of the MSF we here made use of the algorithm for the computation of the entire spectrum of Lyapunov exponents in Ref. \cite{wolf1985determining} (with parameters: integration step size of the Euler algorithm $\delta t=10^{-5}$, length of the simulation $L=2500 s$, windows of averaging $T=0.9L$).

The results are shown in Fig.~\ref{fig:MFS}  for the R\"ossler oscillator and in Fig.~\ref{fig:MFSLorenz} for the Lorenz system. Both cases exhibit a variety of behaviors that actually encompass all possible classes of MSF. In the case of R\"ossler oscillator we have one class III example (Fig.~\ref{fig:MFS}(a)), one class II example (Fig.~\ref{fig:MFS}(e)), while all remaining cases do correspond to class I. In the case of the Lorenz system we have several examples of class I behavior (Fig.~\ref{fig:MFSLorenz}(c), (f), (g) and (h)); three class II examples (Fig.~\ref{fig:MFS}(a),(d) and (e)), and one class III example with a very narrow region for synchronization (Fig.~\ref{fig:MFS}(b)). Moreover, in Fig.~\ref{fig:MFS}(i) the MSF assumes negative values in two different intervals of $\alpha$; overall, this represents a further example of class III behavior, providing however the extra scenario where increasing the coupling strength one can achieve alternating regions of synchronization and desynchronization.

Finally, we apply our method to a real world network modeling the interactions between the members of the Zachary karate club \cite{zachary1977information}. The network consists of $N=34$ nodes and $78$ links; moreover the links form 45 triangles. From this network several simplicial complexes can be constructed, depending on which and how many nodes forming a triangle are effectively taken into consideration as forming part of a 2-simplex or, on the contrary, as determining only three pairwise interactions.

At first, let us consider the case where all triangles are considered as 2-simplexes. We associate to each node a R\"ossler oscillator and focus on the class III case, selecting the coupling functions as $\mathbf{g}^{(1)}(\mathbf{x}_i,\mathbf{x}_j)=[x_j^3-x_i^3,0,0]^T$ and $\mathbf{g}^{(2)}(\mathbf{x}_i,\mathbf{x}_j,\mathbf{x}_k)=[x_j^2x_k-x_i^3,0,0]^T$. With these assumptions, the dynamics of each node $i$ is described by

\begin{equation}
\label{eq:rosslernets}
%\left \{
\begin{array}{l}
\dot{x}_{i} = -y_{i} -z_{i} -\sigma_1\sum\limits_{j=1}^N\mathcal{L}_{ij}^{(1)}x_j^3-\sigma_2\sum\limits_{j=1}^N\sum\limits_{k=1}^N\tau_{ijk}^{(2)}x_j^2x_k,\\
\dot{y}_{i} = x_{i} + a y_{i},  \\
\dot{z}_{i} = b + z_{i} (x_{i} -c).\\
\end{array}
%\right.
\end{equation}

Eqs. (\ref{eq:rosslernets}) are then simulated for different values of  $\sigma_1$ and $\sigma_2$. The average synchronization error and the predictions provided by the MSF \eqref{eq:MSFsimpliciale} are illustrated in Fig.~\ref{fig:Zac}(a) that shows the crucial role played by the pairwise links, as synchronization turns out to be impossible when only three-body interactions are considered, i.e., when $\sigma_1=0$.

Next, we take the original network, and build  different SCs by considering an increasing percentage (labelled as $p_{2s}$) of triangles in the original structure as true 2-simplexes. For each of these structure we determine the effective matrix $\mathcal{M}$ in \eqref{eq:MSFnatural}, and calculate its spectrum of eigenvalues, and in particular we calculate the quantities $\lambda_2(\mathcal{M})$ and $\lambda_2(\mathcal{M})/\lambda_N(\mathcal{M})$. The former quantity provides the scaling of synchronization for class II systems, while the latter quantity ($\lambda_2/\lambda_N$) is a proxy of synchronizability for class III systems. The larger are the two quantities, the easier is to obtain synchronization. Fig.~\ref{fig:Zac}(b) and (c) illustrate the results at three values of $r=\frac{\sigma_2}{\sigma_1}$. One finds that increasing $p_{2s}$ has the effect of increasing $\lambda_2$ (thus it facilitates synchronization in class II systems), but simultaneously dwindles $\lambda_2/\lambda_N$ (thus hindering synchronization in class III). Furthermore, Fig.~\ref{fig:Zac} reveals that a larger value of $r=\frac{\sigma_2}{\sigma_1}$ leads to larger values of $\lambda_2$, but smaller values of $\lambda_2/\lambda_N$, thus suggesting a beneficial impact of stronger three-body interactions for class II systems and an opposite effect on class III systems.

\section{Conclusions}

Collective emergent phenomena in complex systems are the result of the interactions of many elementary systems, that may occur through different mechanisms. We have here formulated the most general model accounting for many-body interactions of arbitrary order among dynamical systems of arbitrary nature, and we have given explicit necessary conditions for synchronization to set up in these structures in a stable way.

Under the only hypothesis of non-invasiveness of the coupling functions (which is the only assumption impossible to be disregarded, as it is the fundamental basis for the very same existence and invariance of the synchronization solution), we have derived the conditions for stability of the synchronous motion, which involve the use of generalized Laplacian matrices mapping the effects of high order interactions.
Our approach ultimately provided a Master Stability Function, which formalizes the interplay between topology of the SC and dynamics of the single units.
Moreover, we have even shown that in some specific cases the structures associated to the interactions of diverse orders assume special forms that further simplify the problem.
Finally, our theoretical derivations have been complemented by a series of numerical results, which have fully confirmed the validity and generality of the approach.

Our results pave the way to several novel studies.

First, the generality of the assumptions made renders it applicable in a wide range of practical cases, and we expect that our method could be of value in a plethora of experimental and/or practical circumstances, in order to make a series of a-priori predictions on the emergence of synchronization.

Second, the fact that our method can be used irrespectively on the coupling functions offers the possibility to apply it for the investigation of diverse coupling mechanisms that may occur at different orders of the interactions. In particular, questions like what exact role do such interactions play in shaping the path to synchronization and its robustness against heterogeneities in the oscillator dynamics,  or what is the difference in using one or another coupling mechanism, can actually be tackled and clarified by our approach. Answering these questions, indeed, is of crucial  importance from the perspective of engineering mechanisms for achieving synchronization in man-made systems. For instance, power grids are currently synchronized by exploiting only pairwise interactions, whereas more functional and more performing configurations could be designed, thanks to our method, by the use of higher order interactions.

Third, our study focuses on what is possibly the most common and widely studied form of synchronization, that is, the regime where all the units follow the same trajectory. However, as also mentioned in the introduction, many other different forms of synchronization exist, including cluster synchronization, chimera and Bellerophon states, remote synchronization, etc... All such states have been so far studied in structures with pairwise interactions. The emergence of such states, or even of novel ones, in SCs, as well as their stability,  are very intriguing problems and certainly constitute directions for further research.

\section*{Acknowledgements}

F.D.P., S.L. and S.B. acknowledge funding from the
project \textit{EXPLICS} granted by the Italian
Ministry of Foreign Affairs and International Cooperation.

L. V. G. and M. F. acknowledge the support of the Univ. of Catania under the framework "Fondi per la ricerca di ateneo - piano per la ricerca 2016/2018".

%\clearpage

%\bibliography{msf}

\end{document}